\documentclass[%
 reprint,
 amsmath,amssymb,
 aps,
]{revtex4-2}

\usepackage{graphicx}
\usepackage{dcolumn}
\usepackage{bm}

\usepackage[T1]{fontenc}
\usepackage[usenames,dvipsnames]{xcolor}
\usepackage{hyperref}
\usepackage[version=3]{mhchem}

\usepackage{threeparttable}
\usepackage{longtable}

\begin{document}


\title{Delving into the Catalytic Mechanism of Molybdenum Cofactors: A Novel Coupled Cluster Study}

\author{Marta Gałyńska}
 \email{marta.galynska@gmail.com}
 \affiliation{Institute of Physics, Faculty of Physics, Astronomy, and Informatics, Nicolaus Copernicus University in Toruń, Grudziądzka 5, 87-100 Toruń, Poland}
\author{Matheus Morato F. de Moraes}%
 \email{matheusmorat@gmail.com}
 \affiliation{Institute of Physics, Faculty of Physics, Astronomy, and Informatics, Nicolaus Copernicus University in Toruń, Grudziądzka 5, 87-100 Toruń, Poland}
\author{Pawe\l{} Tecmer}%
 \affiliation{Institute of Physics, Faculty of Physics, Astronomy, and Informatics, Nicolaus Copernicus University in Toruń, Grudziądzka 5, 87-100 Toruń, Poland}
\author{Katharina Boguslawski}%
 \email{k.boguslawski@fizyka.umk.pl}
 \affiliation{Institute of Physics, Faculty of Physics, Astronomy, and Informatics, Nicolaus Copernicus University in Toruń, Grudziądzka 5, 87-100 Toruń, Poland}


\date{\today}

\begin{abstract}
In this work, we use modern electronic structure methods to model the catalytic mechanism of different variants of the molybdenum cofactor (Moco).
We investigate the dependence of various Moco model systems on structural relaxation and the importance of environmental effects for five critical points along the reaction coordinate with the DMSO and NO$_3^-$ substrates. 
Furthermore, we scrutinize the performance of various coupled-cluster approaches for modeling the relative energies along the investigated reaction paths, focusing on several pair coupled cluster doubles (pCCD) flavors and conventional coupled cluster approximations. 
Moreover, we elucidate the Mo--O bond formation using orbital-based quantum information measures, which highlight the flow of $\sigma_{\rm M-O}$ bond formation and  $\sigma_{\rm N/S-O}$ bond breaking.
Our study shows that pCCD-based models are a viable alternative to conventional methods and offer us unique insights into the bonding situation along a reaction coordinate.
Finally, this work highlights the importance of environmental effects or changes in the core and, consequently, in the model itself to elucidate the change in activity of different Moco variants.
\end{abstract}

\maketitle


\section{Introduction}

The pterin-based molybdenum cofactor (Moco), also known as molybdopterin, contains a mononuclear molybdenum atom coordinated with sulfur atoms from the pterin derivative groups. 
Moco is an essential component in various enzymes, primarily categorized into three families: sulfite oxidase, xanthine oxidase, and dimethylsulphoxide reductase (DMSOR).\cite{kisker1997review, hill1996review}
The DMSOR family is the largest among them and includes various enzymes responsible for nitrogen (periplasmic nitrate reductase -- Nap, trimethylamine N-oxide reductase -- TMAOR), sulfur (dimethyl sulfoxide reductase - DMSOR) or carbon (formate dehydrogenase) metabolism.
Enzymes such as Nap, TMAOR, or DMSOR, catalyze the oxygen-atom transfer on the Mo active site, changing its oxidation state from Mo(IV) to Mo(VI). 
A detailed understanding of the coordination sphere of Moco and its environment within these enzymes is crucial for elucidating their catalytic mechanisms and broader physiological roles in cellular metabolism. 
Despite its widespread presence in many metalloenzymes and its role in catalyzing essential metabolic reactions, the complete functionalization of Moco is yet to be explained. 

Modifying Moco's coordination or altering its surroundings can provide further insights into the catalytic activities of these enzymes. In a recent experimental study by Mintmier and coworkers,\cite{mintmier_moco} the molybdenum-coordinated cysteine of the periplasmic nitrate reductase (Nap) enzyme in \textit{Campylobacter jejuni} was altered to another amino acid to investigate the impact of the Moco residue on the enzyme's catalytic activity. 
They observed that, with the substitution of the native cysteine (Cys, C) with serine (Ser, S) or aspartic acid (Asp, D), the catalytic activity of NO$_3^-$ reduction is preserved.
The D variant of the Nap enzyme was the only one that showed some marginal catalytic activity concerning DMSO, while the native and S variants were inactive in the presence of DMSO.
On the other hand, the Nap variants coordinated with the C and S ligands catalyzed TMAO, while the D one was inactive.
This result is surprising since the TMAOR enzymes are much more selective and catalyze mainly N-oxides, while DMSOR shows activity for both N-oxides and S-oxides and is usually able to reduce TMAOR.~\cite{johnson2001Moco, moula-ic-2013}
The authors concluded that this behavior could be related to differences in the closest surrounding of the Nap and DMSO enzymes, not necessarily the ligand itself.~\cite{mintmier_moco, johnson2001Moco} 
Lim et al.~\cite{Lim2000} successfully reduced DMSO using only a Mo-containing complex with similar coordination but significantly truncated ligands. 

The reaction with molybdopterin was also investigated with quantum chemical methods.~\cite{webster2001Moco, thapper2002Moco, severen2014, ryde2009, hofmann2007, dong2017, li2013, li2014}
Even though, for such systems, DFT calculations strongly depend on the choice of the exchange--correlation functional,~\cite{li2013} DFT calculations are the main driving force for quantum chemical modeling of bioinorganic systems. 
So far, only the model compounds have been studied with coupled-cluster-type methods.~\cite{li2013, mata-jctc-2014}
The considerable molecular size of the realistic Moco-containing systems limits the routine application of more reliable wave function-based quantum chemical methods.
To bridge the gap between the large system size and reliable description of electron correlation effects, we explore the perspective of using the recently developed pCCD-based methods~\cite{tecmer2022geminal} for modeling electronic structures and properties of Moco-based systems.

The pCCD ansatz~\cite{limacher2013new, boguslawski2014efficient, stein2014seniority} was initially introduced as a geminal-inspired wave function model~\cite{tecmer2022geminal} using two-electron functions as building blocks.~\cite{hurley1953molecular, parr1956generalized, bardeen1957theory, parks1958theory, coleman1965structure, miller1968electron, surjan1999introduction, surjan2012strongly, tecmer2014assessing, johnson2013size, johnson2017strategies, fecteau2020reduced, johnson2020richardson, johnson2022bivariational, faribault2022reduced, fecteau2022near, moisset2022density, tecmer2022geminal}
The pCCD molecular orbitals undergo variational optimization to recover size-consistency, ~\cite{boguslawski2014efficient, limacher2014influence, boguslawski2014nonvariational, boguslawski2014projected, stein2014seniority} which results in localized, symmetry-broken orbitals, facilitating the simulation of quantum states with (quasi-)degeneracies.~\cite{boguslawski2016analysis}
The resulting orbital optimized (oo)-pCCD molecular basis is not restricted by the active space size as for standard multireference methods and, therefore, provides an efficient treatment of electron correlation effects. 
Numerical examples include bond-breaking processes in small molecules,~\cite{tecmer2014assessing, limacher2015orbital, tecmer2015singlet, brzek2019benchmarking, henderson2019geminal, nowak2021orbital, leszczyk2021assessing, leszczyk2022, pccd-dipole-moments-2024} compounds containing heavy elements like lanthanides~\cite{tecmer2019modeling} and actinides,~\cite{tecmer2015singlet, garza2015actinide, boguslawski2016targeting, boguslawski2017erratum, nowak2019assessing, leszczyk2022, Nowak2023, chakraborty2023} organic electronics,~\cite{jahani2023relationship, tecmer2023jpcl} electronically excited~\cite{boguslawski2016targeting, boguslawski2017erratum, boguslawski2018targeting, nowak2019assessing, kossoski2021excited, bartlett-pccd-tcc} and ionized states.~\cite{boguslawski2021open, mamache2023, tailored-ip-pccd-jctc-2024}

Despite the promising performance of pCCD in capturing static/nondynamic electron correlation effects,~\cite{sinanoglu1963, bartlett_1994,  entanglement_letter} a considerable fraction of correlation energy remains unaccounted for by electron-pair states alone.
This missing correlation energy is commonly associated with broken-pair states, necessitating their incorporation in post-optimizations through various state-of-the-art techniques.~\cite{tecmer2022geminal}
One approach to address dynamical correlation involves a coupled cluster correction on top of the pCCD reference wave function.~\cite{henderson2014seniority, boguslawski2015linearized, leszczyk2021assessing}
The role of the second cluster operator is to incorporate electron excitations beyond electron pairs.
Such a family of methods was introduced as frozen-pair CC (fpCC) and can be viewed as a conventional tailored coupled cluster method.~\cite{kinoshita2005, tailoredcc2006, lyakh2011, tailoredcc2012}
The fpCC ansatz can be simplified even further, taking only the linear terms until the second term of the Baker–-Campbell–-Hausdorff expansion,~\cite{boguslawski2015linearized,boguslawski2017benchmark} that is neglecting all non-linear term associated with broken-pair excitations, resulting in a frozen-pair linearized CC approximation (fpLCC{, originally labeled as pCCD-LCC}). 

An additional advantage of pCCD is the immediate access to their 1- and 2-particle reduced density matrices from which orbital-based entanglement and correlation measures can be derived.~\cite{rissler2006,  barcza2014entanglement, ijqc-2015, ijqc-erratum, nowak2021orbital, Ding2020, qit-concepts-schilling-jctc-2021}
Specifically, these can be used to calculate the single-orbital entropy and orbital-pair mutual information.~\cite{rissler2006, ors_ijqc, barcza_11, entanglement_letter, entanglement_bonding_2013}
While the single-orbital entropy describes the entanglement between orbitals, the mutual information, which includes classical and quantum effects,~\cite{qit-concepts-schilling-jctc-2021, ding2022quantum} serves a measure of electron correlation effects of a quantum system.~\cite{entanglement_letter, ijqc-2015, ijqc-erratum} 
These measures can be used, for instance, to monitor bond-breaking/forming processes,~\cite{entanglement_bonding_2013, mottet, corinne_2015, zhao2015, tecmer2015singlet, leszczyk2022, roland-runo} determine bond-orders,~\cite{entanglement_bonding_2013, mottet} and guide the choice of optimal active spaces in multireference calculations.~\cite{ijqc-2015, stein2016, boguslawski2017}

In this study, we aim to evaluate the performance of various pCCD-based flavors for modeling reactions of the Moco system coordinated with various amino acid ligands with DMSO and NO$_3^-$. 
Their reliability is compared with conventional CC approaches.
Finally, a pCCD-based entanglement and correlation analysis will shed more light on orbital interactions in various Moco-coordinated systems and the change in chemical bonding along the investigated Moco series.  


\section{Computational details}\label{sec:comput_details}
\subsection{Methods and basis sets}\label{sec:comput_details-methods}
All structures were relaxed using the BP86 exchange--correlation functional.~\cite{perdew86, becke88}
We chose the Jorge-ZORA-TZP~\cite{basis_set_zora_jorge} basis set for the Mo atom and the Karlsruhe-type basis set optimized for ZORA (ZORA-def2-TZVP)~\cite{def2-tvzp,zora_pantazis} and auxiliary basis set SARC/J\cite{zora_pantazis} for the other atoms.
The conductor-like polarizable continuum model~\cite{pcm-chem-rev-2005} (CPCM) with a dielectric constant 4 ($\epsilon$ = 4) was incorporated to simulate the protein environment.
{No additional dispersion correction was applied to avoid double-counting of dispersion effects incorporated in the CPCM model.}
All DFT calculations were conducted using the Orca 5.0.4 quantum chemistry software.~\cite{orca-2022}

Subsequently, single-point calculations were performed on these optimized structures using different coupled-cluster flavors as implemented in the PyBEST v1.4.0dev0 software package.~\cite{pybest-paper, pybest-paper-update1-cpc-2024, pybest1.3.1-zenodo, pybest-web} 
These included orbital-optimized pair coupled-cluster doubles (oo-pCCD), which served as a reference wave function for frozen-pair CCD (fpCCD) and frozen-pair linearized CCD (fpLCCD) calculations.
{In the following, we will drop the ``oo" prefix in the acronym and use the common abbreviation pCCD to indicate the (variationally) orbital-optimized pCCD variant.
Note, however, that all pCCD-based calculations are done in the optimized natural pCCD molecular orbital basis.}
Conventional CCD and CCSD calculations using canonical Hartree-Fock orbitals were added for comparison.

Furthermore, domain-based local pair natural orbital (DLPNO) CCSD~\cite{dlpno_neese_1} and CCSD(T)~\cite{dlpno_neese_2} energies were calculated with and without the CPCM model in Orca 5.0.4~\cite{orca-2022} to examine the importance of triple excitations and the effect of solvent on the energies of critical points along the reaction energy paths.
{Specifically, the CPCM model is implemented as a perturbation theory energy (PTE) scheme, where its contribution to the CC energy occurs through the term $\frac{1}{2} Q_0 \cdot V_0$ and implicitly through the Fock matrix elements $F_{ia}$ ($F_{ia}$ = $F_0 + Q_0 \cdot v_{ia}$).}~\cite{cammi2009quantum,caricato2011ccsd}
All CC calculations utilized the Jorge-DKH2-DZP basis set~\cite{basis_set_dkh_1, basis_set_dkh_2} and a frozen core approximation (one orbital for C, N, O, five orbitals for S, and eighteen orbitals for Mo were kept frozen).
Scalar relativistic effects were accounted for using the second-order Douglas--Kroll--Hess (DKH2) Hamiltonian.~\cite{dkh1, dkh2, reiher_book, tecmer2016}
{We should note that the choice of different scalar relativistic Hamiltonians does not affect the valence properties of the investigated systems as long as the basis set is optimized for a specific Hamiltonian.\mbox{\cite{pawel1,tecmer2016}}}

{Furthermore, since the chosen Mo-based compounds do not approach the strong correlation regime, CCSD(T) is a valid method of choice to generate theoretical reference data. To decrease the computational cost of conventional CCSD(T) calculations, we applied the DLPNO variant instead.
Previous studies also indicate that pCCD-based methods provide smaller errors with respect to highly accurate wavefunction-based methods (like DMRG or CCSD(T) or CR-CCSD(T)) than conventional CC methods of similar scaling, namely CCSD.\mbox{\cite{boguslawski2017benchmark,boguslawski2018targeting,nowak2019assessing,leszczyk2021assessing,boguslawski2021open}}
This study will, hence, allow us to assess the performance of post-pCCD approaches for transition metal chemistry for the first time.
Finally, we aim to compare pCCD-based methods to other CC flavors, like conventional CCSD and DLPNO-CCSD(T) within a given basis set, while the orbital entanglement and correlation measures are less sensitive to the basis set size as long as we consider moderate to strong orbital-pair correlations.
Since we further focus on relative energies, the basis set size will, therefore, have a smaller effect on our numerical results and the final conclusions (see, for instance, refs\mbox{\citenum{tecmer2014assessing,boguslawski2017benchmark,boguslawski2015linearized,leszczyk2021assessing, pccd-dipole-moments-2024}} for the basis set dependence of pCCD-based methods). 
This validates the choice of a DZP-quality basis set in our CC calculations.}
\begin{figure}[t]
\centering
  \includegraphics[width=9cm]{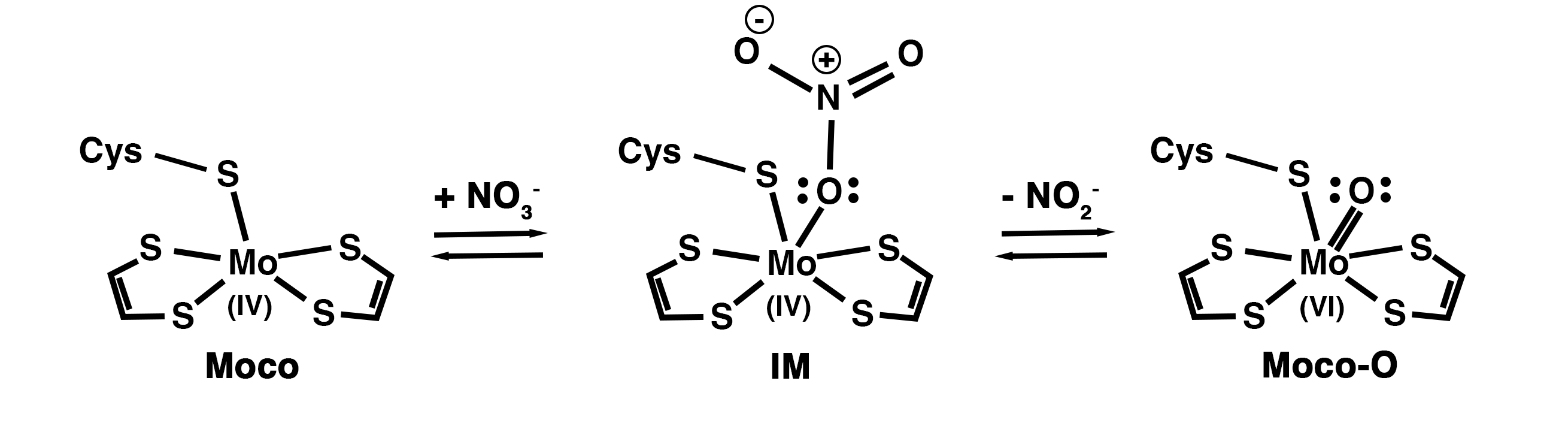}
  \caption{{The catalytic reaction between the molybdenum cofactor (Moco) and NO$_3^-$ in the NapA protein follows a two-step mechanism. In the first step, an intermediate state (IM) is formed by creating a Mo--O bond. In the second step, the NO$_2^-$ molecule dissociates, forming an oxo group and changing the oxidation state of Mo from IV to VI. Note that TS1 and TS2 are not shown in the Figure.}}
  \label{fgr:mechanism}
\end{figure}

\begin{figure*}
 \centering
 \includegraphics[height=13cm]{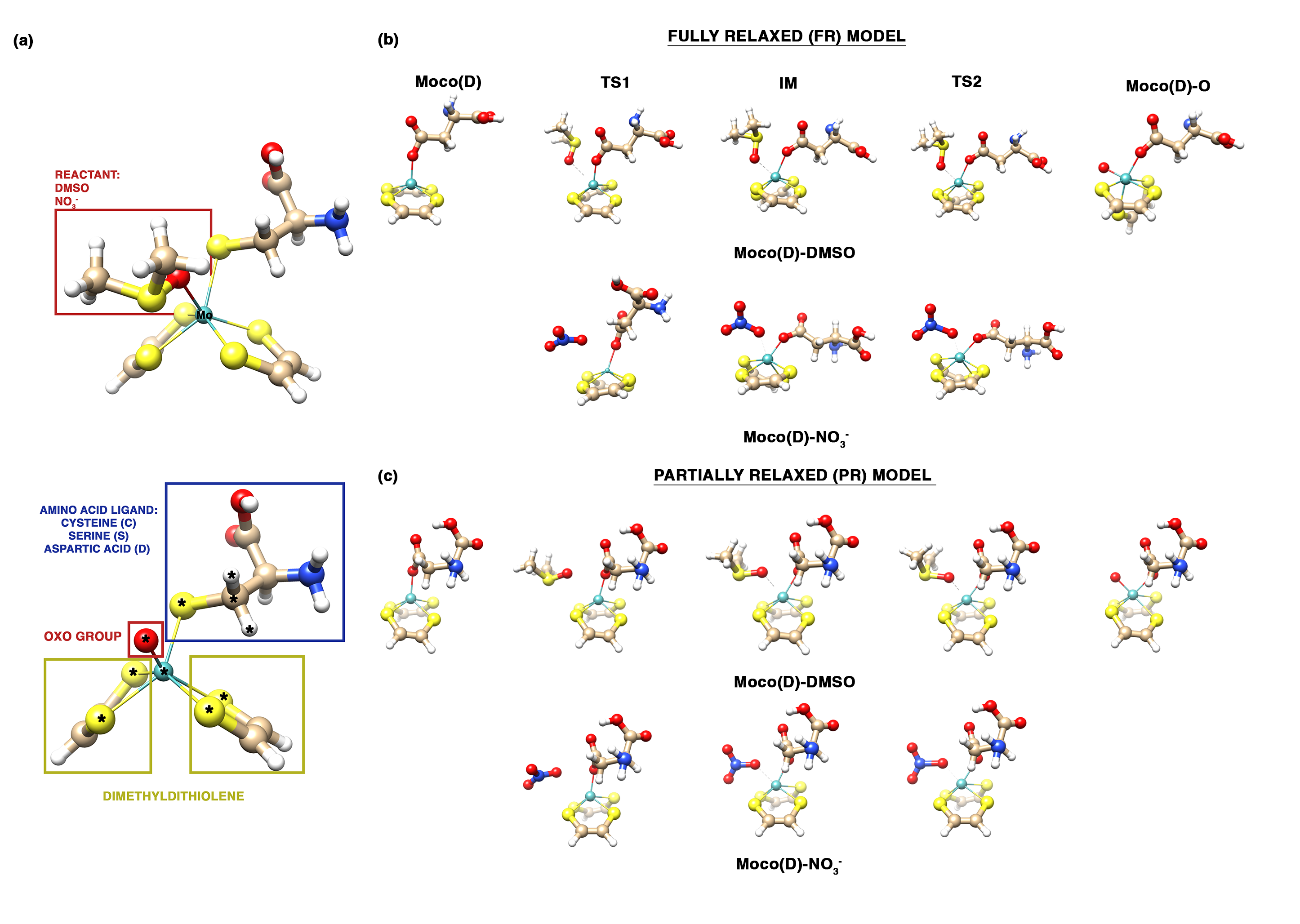}
 \caption{(a) Moco's intermediate state and oxidized structure were modeled with two dimethyldithiolene units, an amino acid ligand (C, S, and D), and an oxo group. The stars indicate atoms, which were relaxed in a partially relaxed model ({PR}), see subsection~\ref{sec:comput_details-structures} for more details. {Five critical points along the reaction path of Moco(D) with DMSO and NO$_3^-$ using two relaxation schemes: (b) fully relaxed (FR) and (b) partially relaxed (PR).}}
 \label{fig:structure}
\end{figure*}
\subsection{Generation of model structures}\label{sec:comput_details-structures}
The initial geometry was derived from the crystallographic structure of the periplasmic nitrate reductase (Nap) from \textit{Escherichia coli}, obtained from the Protein Data Bank~\cite{pdb} (PDB code: 2NYA). 
{The mechanism of the catalytic reaction between Moco and NO$_3^-$ in the native Nap protein involves two steps as shown in Fig. \mbox{\ref{fgr:mechanism}}}
The molybdenum cofactor (Moco) consists of a Mo(IV) atom coordinated with two pyranopterin ligands, with the fifth coordination site provided by the thiolate end of the cysteine (C) amino acid.
Upon oxidation, an additional oxo ligand coordinates to the Mo atom, resulting in a change of the Mo oxidation state from Mo(IV) to Mo(VI).~\cite{mintmier_moco} 

The models used in this study to represent the Moco system are shown in Fig.~\ref{fig:structure}(a). 
The two pyranopterin ligands were approximated by two bidentate dithiolene units coordinated to the molybdenum atom, Mo(S$_2$C$_2$H$_2$)$_2$, while the fifth ligand was represented by an entire amino acid molecule (Cys=HOOCCH(NH$_2$)CH$_2$S$^-$), resulting in a Mo complex ([Mo(S$_2$C$_2$H$_2$)$_2$Cys]$^{-}$).
This representation differs from previous studies, where the amino acid ligand was approximated solely by CH$_3$O$^-$ for serine, etc.\cite{webster2001Moco,thapper2002Moco, dong2017, li2013, li2014} 
Since the C, S, and D amino acids differ only in functional groups, we substituted --S$^-$ by --O$^-$ and --COO$^-$, respectively, to create the [Mo(S$_2$C$_2$H$_2$)$_2$Ser]$^{-}$ and [Mo(S$_2$C$_2$H$_2$)$_2$Asp]$^{-}$ variants. 
For simplicity, the substrates of the different variants of Moco are referred to as Moco(C), Moco(D), and Moco(S), while the oxidized products are denoted as Moco-O(C), Moco-O(D), and Moco-O(S) for cysteine, aspartic acid, and serine, respectively.

Subsequently, the NO$_3^-$ and DMSO molecules were added to create the intermediate (IM) and two transition states: TS1 (between the Moco and intermediate state) and TS2 (between the intermediate state and product{; see also Fig. \mbox{\ref{fgr:mechanism}}}).
{In this study, we followed the computational procedure outlined in \mbox{\citenum{ryde2009,li2014,dong2017}}, where the molecular structures for each point along the reaction coordinate are fully relaxed. However, the final molecular structures turned out to be distorted compared to the structure data available in the Protein Data Bank.\mbox{\cite{pdb}}
To address this distortion problem, we designed two additional models where different groups of atoms are partially fixed.
To that end,} we considered three relaxation schemes: fully relaxed {(FR)}, partially relaxed {(PR)}, and {unrelaxed (UR)}.
In contrast to the {FR} model, where all the atoms are relaxed, the {UR} model only relaxed the {substrate}, all hydrogen atoms, {and the oxo group in the product}, while the rest remained constrained. 
In the {PR} model, the atoms marked with stars in Figure~\ref{fig:structure}(a) were additionally relaxed compared to {UR}. 
Figure~\ref{fig:structure}(b)  depicts the {FR} and {PR} model structures for critical points along the reaction path between Moco(D) and DMSO, and Moco(D) and NO$_3^-$.

In the {FR} model, the nature of critical points was confirmed by harmonic frequency analyses, showing only real frequencies for Moco, IM, and oxidized Moco and one negative for the saddle point structures of TS1 and TS2. 
However, in the {PR} and {UF} models, the substrate, intermediate state, and product exploit the coordinates directly obtained from the crystallographic Nap structure.
Consequently, they do not correspond to local minima and have a few negative frequencies.
The two transition states were estimated by stretching the Mo--NO$_3^-$ bond for TS1 and the O--NO$_2^-$ bond for TS2 to approximately the same distance found in the TS1 and TS2 structures of the {FR} model.
Specifically, we manually scanned the potential energy surface around the {FR} distances and selected the geometry corresponding to the highest energy {through constrained geometry optimization, where the Mo--N bond/angles were fixed, while the remaining atoms were allowed to relax freely depending on the chosen model (fully relaxed or partially relaxed)}.
An analogous procedure was used for the DMSO substrate{, where we scanned the proximate PES for different Mo--S bond distances}.
\section{Results and discussion}\label{sec:results}
\subsection{Reaction dependence on structural relaxation}\label{sec:results-energies}
\begin{figure}[ht!]
\centering
  \includegraphics[width=7cm]{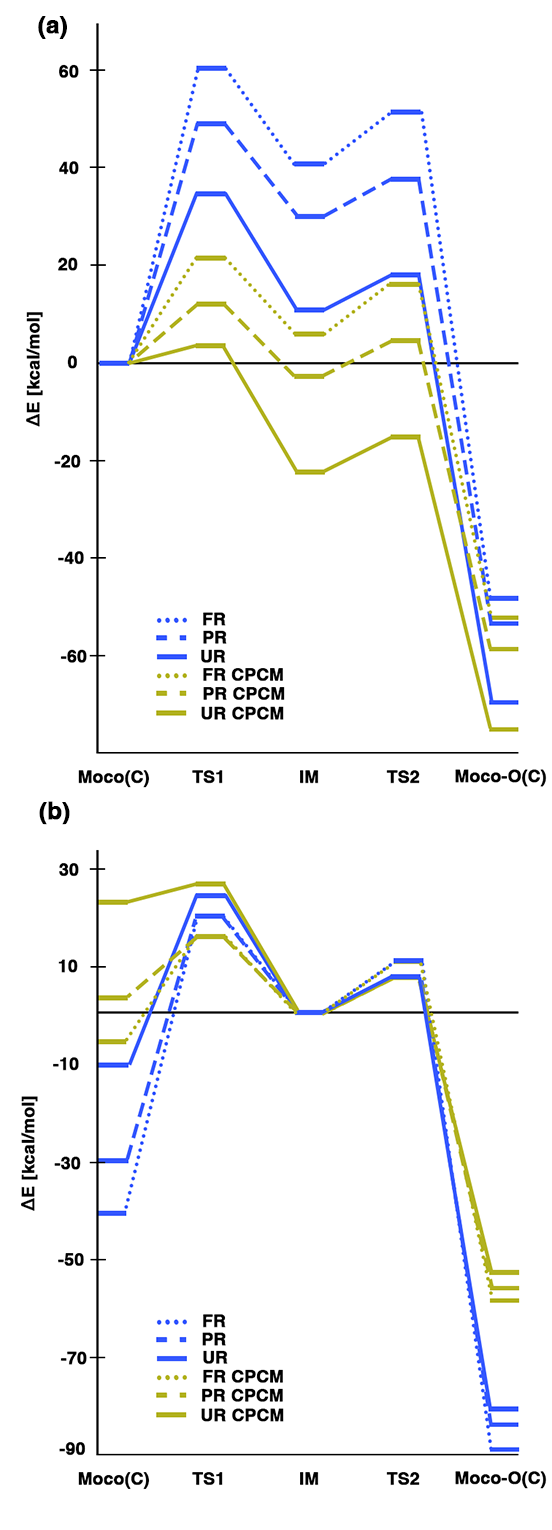}
  \caption{The relative energies [kcal/mol] for five critical points along the reaction path between Moco(C) and NO$_3^-$ calculated using the DLPNO-CCSD/DKH-pVDZ method with three relaxation schemes: {FR} (dotted), {PR} (dashed), and {UR} (solid), both with (marked in {olive}) and without (marked in blue) CPCM. The relative energies were calculated with respect to the (a) Moco(C) and (b) IM energies.}
  \label{fgr:models}
\end{figure}

First, we scrutinize the influence of the relaxation scheme on the reaction path between Moco(C) and NO$_3^-$.
Specifically, we focus on the relative energies of the critical points of the reaction coordinate obtained with DLPNO-CCSD/DKH-pVDZ with and without the CPCM model and all three Moco relaxation models (fully relaxed, partially relaxed, and fully frozen).
The corresponding results are shown in Figure~\ref{fgr:models} and plotted against the (a) Moco(C) and (b) IM reference values.
The relative energies show strong sensitivity to the relaxation schemes when the Moco(C) structures are used as a reference. 
Overall, the relative energies gradually decrease with a decreasing number of relaxed atoms in the model, as shown in Figure~\ref{fgr:models}(a).
Consequently, the highest energy barriers were obtained using the {FR} model, while the smallest ones were obtained using the {UR} model.
Specifically, the relative energies for TS1, IM, and TS2 from the {PR} and {UR} models are lowered by 10.8–13.5 kcal/mol and 25.7–33.2 kcal/mol, respectively, compared to the {FR} model. 
We should stress that the Moco(C)-O values are only reduced by 5.3 kcal/mol for {PR}.
This value increases to 21.5 kcal/mol for {UR}. 
In general, the presence of solvent (more precisely, the CPCM solvation model) further decreases the relative energies by 30.8–38.9 kcal/mol for TS1, IM, and TS2 but only by 4.1–5.3 kcal/mol for the Moco(C)-O structures. 

Interestingly, choosing the IM structures as reference point (Fig.~\ref{fgr:models}(b)) alleviates some of the differences in relative energies.
Specifically, the discrepancies between different relaxation schemes are primarily due to the total energy values of the Moco(C) educt, yielding a range of relative energies from \mbox{-40.9} to 22.2 kcal/mol.
On the other hand, the second barrier (E$_{\rm TS2}-$E$_{\rm IM}$) seems to be independent of the applied relaxation model or whether the calculations are done in vacuo or solution, as these values differ by at most 3.4 kcal/mol. 
Similarly, a slightly larger variation can be observed with respect to TS1, with a maximum difference of 11.0 kcal/mol. 
The relative energies of the oxidized Moco(C) structures fall into two ranges: from -80.5 to -89.0 kcal/mol and from -52.7 to -58.2 kcal/mol when calculated in vacuo and with CPCM, respectively.

These results do not unambiguously substantiate which relaxation scheme is the best for a quantum chemical modeling of the catalytic reaction of Moco's active site. 
To that end, the remaining results will be compared using the {FR} and {PR} models.
\subsection{Relative energies from different CC flavors}\label{sec:results-relative-energies}
\begin{figure}
       \centering
        \includegraphics[height=8.3cm]{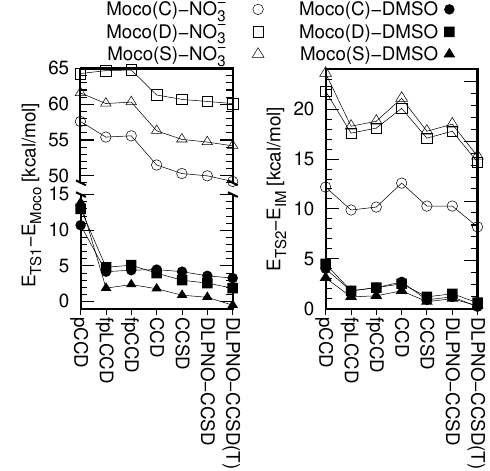} 
        \caption{The first energy barrier (E$_{\rm TS1}-$E$_{\rm Moco}$) and the second energy barrier (E$_{\rm TS2}-$E$_{\rm IM}$) for the reaction paths between different variants of Moco (Moco(C), Moco(D), and Moco(S)) and two {substrates} (DMSO and NO$_3^-$) relaxed with the {FR} model and calculated using various CC methods. These methods include pCCD, fpLCCD, and fpCCD, utilizing the natural pCCD-optimized orbitals, as well as the canonical CCD and CCSD methods, and the DLPNO-CCSD and DLPNO-CCSD(T) variants with Hartee--Fock orbitals used as reference.
        }
        \label{fig:barrier}
\end{figure}

The single-point calculations employing various CC flavors are performed for the critical points along the reaction paths between different variants of Moco and DMSO or NO$_3^-$ derived from the \mbox{{FR}} and {PR} models. 
These CC methods encompass pCCD-based approaches utilizing natural pCCD-optimized orbitals as reference wave functions (pCCD, fpLCCD, and fpCCD), which we compare to the canonical CCD and CCSD methods. 
Furthermore, the impact of triple excitations is examined by comparing the DLPNO-CCSD to the DLPNO-CCSD(T) approximation. 
Additionally, the influence of the protein environment, including the CPCM model on top of DLPNO-CCSD and DLPNO-CCSD(T), is also explored. 
The relative energies obtained with all these methods are summarized in Table~\ref{tbl:energies}, while a graphical representation of the change in barriers with respect to the employed CC method is illustrated in Figure~\ref{fig:barrier}, again with respect to 2 reference poinst, E$_{\rm Moco}$ and E$_{\rm IM}$.
\begin{table*}
\small
  \caption{Relative energies [kcal/mol] of the Moco(C), Moco(D), and Moco(S) structures along the investigated reaction paths involving the NO$_3^-$ and DMSO molecules.
The structures were optimized using two relaxation schemes ({FR} and {PR}), and the final energies were calculated using canonical CCD and CCSD, DLPNO-CCSD, and DLPNO-CCSD(T) methods with and without CPCM. The orbital-optimized pCCD method was used as reference wave function for fpLCCD and fpCCD. 
}
  \label{tbl:energies}
  \begin{tabular*}{\textwidth}{@{\extracolsep{\fill}}l  r r r r r r r r r}
    \hline
     & pCCD  & fpLCCD & fpCCD & CCD & CCSD & DLPNO- & DLPNO-   & DLPNO-        & DLPNO-  \\
           &         &        &     &      & & CCSD   & CCSD(T)  & CCSD {CPCM}  &CCSD(T) {CPCM} \\
    \hline
    \multicolumn{10}{c}{\textbf{FR Moco(C)--DMSO}}\\
    Moco(C)   &  0.0  &  0.0  &   0.0	&  0.0  &  0.0  &  0.0 &  0.0  &   0.0 &   0.0  \\
    TS1       &  10.7 &  4.2  &   4.4	&  4.5  &  4.2  &  3.6 &  3.3  &   4.6 &   4.4  \\
    IM        &  6.4  & -6.6  &  -6.1	& -6.5  & -6.7  & -7.1 & -7.7  &  -9.5 & -10.1  \\
    TS2       & 10.4  & -4.8  &  -4.0	& -3.8  & -5.8  & -5.9 & -7.5  &  -7.7 & -8.9  \\
    Moco(C)-O  & -78.6 & -87.6 &	-93.6	& -80.9	& -77.8	& -78.4& -80.8 & -73.7 & -76.4  \\
    \multicolumn{10}{c}{\textbf{FR Moco(D)--DMSO}}\\ 
    Moco(D)   & 0.0   &	 0.0	&    0.0 &	 0.0  &   0.0 &	0.0	 &   0.0 &	0.0	  &   0.0 \\
    TS1        & 13.0  &	 4.8	&    5.1 &	 4.0  &   3.0 &	 2.6 &	 1.9 &	1.9	  &   1.1 \\ 
    IM        & 10.5  & -5.6    &	-4.9 &	-6.5  &	 -8.0 &	-8.3 &	-9.8 &	-10.2 &	-11.7 \\
    TS2       & 15.0  & -3.8    &	-2.8 &	-4.0  &	 -6.8 &	-6.8 &	-9.2 & 	-8.2  &	-10.3 \\
    Moco(D)-O  & -68.3  & -81.3   &  -87.3 & -75.4  &	-73.1 &-73.7 &	-77.1 &	-69.0 &	-72.8 \\
    \multicolumn{10}{c}{\textbf{FR Moco(S)--DMSO}}\\
    Moco(S)   & 0.0   &	 0.00	&  0.0  &	 0.0  &	 0.0  &  0.0  &	 0.0	&  0.0 &  0.0\\
    TS1        & 13.9  &	 1.9	&  2.4  &	 1.8  &	 0.9  &  0.6  & -0.5    &  3.1 &  1.9\\
    IM         & 12.7  &	-4.5	& -3.7  &	-4.3  &	-5.8  & -5.9  & -7.8	& -6.1 & -7.8\\
    TS2        & 15.8  &	-3.3	& -2.4  &	-2.5  &	-5.1  & -4.9  &  -7.5   & -4.8 & -7.0\\
    Moco(S)-O  & -76.0  &	-88.0	& -94.0 &	-81.5 &	-79.2 & -78.4 &	-80.8   &-73.7 & -76.4\\
    \hline   
    \multicolumn{10}{c}{\textbf{FR Moco(C)--NO$_3^-$}}\\
    Moco(C) & 0.0 &	0.0	  &  0.0 &	0.0	  &   0.0 &	0.0  &	0.0  &	0.0	 &  0.0 \\
    TS1      & 64.3 &	64.7  &	64.8 &	61.3  &	60.7  &	60.4 &	60.1 &	21.5 &	21.3 \\
    IM       & 51.8 &	43.3  &	 43.7 &	42.4  &	41.5 &	40.9 &	39.8 &	6.0	 &    5.0 \\
    TS2      & 64.0 &	53.2  &	53.9  &	55.0  &	51.8 &	51.2 &	48.0 &	16.1 &	12.9 \\
    Moco(C)-O   &-53.8 &	-54.5 &	-54.9 &	-49.0 &	-47.7 &	-48.1 &	-50.4 &	-52.2 &	-54.4 \\
    \multicolumn{10}{c}{\textbf{FR Moco(D)--NO$_3^-$}}\\
    Moco(D) & 0.0	&0.0   &	 0.0 &	0.0	  &   0.0 &	0.0	  &  0.0  &	0.0	 &   0.0 \\
    TS1      & 57.6	& 55.4 &	55.6 &	51.5  &	50.3  &	50.0 &	49.2  &	14.7 &	14.0  \\
    IM       & 57.3	& 47.8 &	48.2 &	45.8  &	43.9  &	43.3 &	41.6  & 9.6	 &   8.0 \\
    TS2      & 79.1	& 65.4 &	66.3 &	65.9  &	61.0  &	61.1 &	56.3  &	26.8 &	22.2 \\
    Moco(D)-O  & -43.4 &	-48.3 &	-48.6 &	-43.5 &	 -43.0 & -43.4 &-46.7 &	-47.5 &	-50.7 \\
    \multicolumn{10}{c}{\textbf{FR Moco(S)--NO$_3^-$}}\\
    Moco(S)    & 0.0	& 0.0	 &  0.0	  &   0.0	& 0.0	&0.0	& 0.0	& 0.0	 &   0.0 \\
    TS1        & 61.6	& 60.1	 & 60.3	  &  56.3	& 55.1	&54.7   &	54.2 &	17.6 &	17.1 \\
    IM         & 64.3	& 56.5   &	56.9  &	55.3	& 53.5	& 53.2	& 51.5	 & 17.1  & 	15.7 \\
    TS2        & 87.9	& 74.8   &	75.7  &	76.4	& 71.3	& 71.7	& 66.8	& 36.2	 &   31.3 \\
    Moco(C)-O  & -51.2& -54.9 &	-55.3 &	-49.6	& -49.1	& -49.4	& -52.4	& -49.8	 &   -52.7 \\
    \hline   
    \multicolumn{10}{c}{\textbf{PR Moco(C)--DMSO}}\\
    Moco(C)   & 0.0 &	0.0   &	  0.0 &	  0.0 &	  0.0 &	  0.0 &	  0.0 &   0.0 &	  0.0 \\
    TS1       & 7.2 &	0.7	  &   0.9 &	  0.4 &	 -0.1 &	 -0.7 &	 -0.8 &	 -2.0 &	 -2.2 \\
    IM         &-0.5 &	-10.7 &	-10.5 &	-10.7 &	-10.7 &	-11.6 &	-11.5 &	-17.8 &	-17.8 \\
    TS2       & 5.8 &	 -6.1 &	 -5.6 &	 -4.4 &	 -6.0 &	 -6.2 &	 -7.1 &	-14.0 &	-14.5 \\
    Moco(C)-O  & -86.8 &	-91.6 &	-97.8 &	-85.9 &	-82.8 &	-83.6 &	-85.7 &	-80.1 &	-82.4 \\
    \multicolumn{10}{c}{\textbf{PR Moco(D)--DMSO}}\\
    Moco(D)  & 0.0 &   0.0 &  0.0 &	   0.0 &   0.0 &   0.0 &    0.0 &  0.0 &	0.0 \\
    TS1      & 8.4 &   0.8 &  1.0 &	   0.1 &  -0.7 &  -1.4 &   -1.9	& -2.9 &	-3.6 \\
    IM       & 3.5 &  -9.6 & -9.2 &	  -10.5	& -11.1 & -12.0 & -12.7	& -18.4 &	-19.3 \\
    TS2      & 9.9 &  -4.9 & -4.2 &	  -3.9	& -6.3	& -6.3	& -8.0	& -14.3	&   -15.7 \\
    Moco(D)-O & -78.4 & -86.5 & -92.5 &  -80.8	& -78.3	& -79.6	& -82.8	& -76.7	&   -80.2 \\
    \hline
    \multicolumn{10}{c}{\textbf{PR Moco(C)--NO$_3^-$}}\\
    Moco(C)  &  0.0  & 0.0    &	 0.0  &	0.0	 &    0.0 &	  0.0 &	0.0   &	0.0 &	0.0 \\
    TS1       & {5}4.5  &	52.0   &	52.1  & 50.6 &	49.8  &	49.1  &	 48.9 &	12.2 &	12.1 \\
    IM       & 38.7  &	32.2   &	32.4  &	31.9 &	31.2  &	 30.1 &	29.6  &	-2.8 &	-3.3 \\ 
    TS2       & 48.0  &  39.3  &	39.8  &	41.0 &	38.6  &	 37.7 &	35.6  &	4.6	 &  2.4  \\
    Moco(C)-O & -61.9 &	-58.5  &   -59.2  &-54.0 & -52.7  &	-53.4 &	-55.3 &	-58.6 & -60.4 \\
    \multicolumn{10}{c}{\textbf{PR Moco(D)--NO$_3^-$}}\\
    Moco(D)   & 0.0  &	0.0   &	  0.0 &	0.0   &	0.0 &	0.0 &	0.0	& 0.0	&0.0 \\
    TS1       &  52.9 &	51.3  &  51.4 & 47.3  & 46.3	&  45.0	& 44.3  &	9.7	&  9.0 \\
    IM        &  43.6 &	32.6  &	 33.1 &	31.4  &	30.1 &	29.2 &	27.7 &	-3.8 &	 -5.3 \\
    TS2       &   58.2 &	43.9  &  44.8 &	45.4  &	41.7 &	40.9 &	36.9 &	7.5	 & 3.5 \\
    Moco(D)-O &  -53.5 &-53.4 &	-53.8 &	-48.9 &	-48.2 &	-49.4 &	-52.4 &	-55.2 &	-58.2 \\
    \hline
 \end{tabular*}
\end{table*}

In general, among all the methods that account for dynamical correlation (fp-pLCCD, fp-CCD, CCD, CCSD, DLPNO-CCSD, and DLPNO-CCSD(T)), the lowest energies for TS1, IM, and TS2 are obtained from DLPNO-CCSD(T) for the DMSO reaction with different variants of Moco.
However, the differences between DLPNO-CCSD(T) and the other listed methods vary, with discrepancies of up to 5.1 kcal/mol for {FR}-Moco(S)-DMSO, 3.8 (2.7) kcal/mol for {FR}-Moco(C)-DMSO ({PR}-Moco(C)-DMSO), and 6.4 (4.2) kcal/mol for {FR}-Moco(D) ({PR}-Moco(D)), respectively.
We observe a similar trend, though with slightly larger differences, for the reaction with NO$_3^-$, with discrepancies of up to 9.6 kcal/mol for {FR}-Moco(S)-NO$_3^-$, 7.0 (5.5) kcal/mol for {FR}-Moco(C)-NO$_3^-$ ({PR}-Moco(C)-NO$_3^-$O), and 10.0 (8.5) kcal/mol for {FR}-Moco(D)-DMSO ({PR}-Moco(D)-DMSO).
Noteworthy, the structures obtained from the {PR} model exhibit slightly smaller differences between these methods.
As expected, pCCD, the simplest model studied, yields the largest differences, shifting the relative energies of TS1, IM, and TS2 upwards with respect to DLPNO-CCSD(T), ranging from 6.4 to 24.3 kcal/mol for the DMSO reaction and from 5.5 to 21.3 kcal/mol for the NO$_3^-$ reaction, respectively.

While pCCD shows a significant upward shift in relative energies with respect to DLPNO-CCSD(T) for TS1, IM, and TS2, the relative energies of the oxidized Moco-O structure do not change as much, generally being lower (except for Moco(D)-O) than DLPNO-CCSD(T). 
A similar trend considering the oxidized Moco-O product is observed for fpLCCD and fpCCD, yielding lower energies than DLPNO-CCSD(T), possibly due to the different description of the reference wave function, particularly the natural pCCD-optimized orbitals for the Mo complex, where the Mo atom is in the VI rather than IV oxidation state.

The addition of the solvent model has a tremendous effect on the relative energies of TS1, IM, and TS2 along the Moco-NO$_3^-$ reaction pathways, resulting in a downward shift of 32.9-37.1 kcal/mol when considering both DLPNO-CCSD and DLPNO-CCSD(T). 
Although the CPCM inclusion changes drastically the binding energy (i.e., E$_{\rm TS1}-$E$_{\rm Moco}$), its effect over the reaction barrier is minor (i.e., E$_{\rm TS2}-$E$_{\rm IM}$).
Therefore, the gap between the barriers determined for selected catalytic steps and sets of ligands is essentially constant with respect to the computational methodology employed.
Numerically, the turnover energetic barriers increase by around 7.5 and 8.5 kcal/mol, respectively, going from Moco(C)-NO$_3^-$ to Moco(D)-NO$_3^-$ to Moco(S)-NO$_3^-$.
Such a change in barrier height agrees qualitatively with the experimentally observed increase in activation free energy, where $\Delta G^\ddag$ increases by 2.9 and 1.7 kcal/mol with the exchange of the cysteine residue by serine and aspartate residues, respectively.
On the contrary, the use of CPCM does not shift the reaction barriers involving DMSO, where the relative energies decrease by up to 8 kcal/mol. 
The relative energies of the oxidized-Moco were lowered by at most 5.9 kcal/mol for the reactions involving NO$_3^-$, while they exhibited an upward shift ranging between -3.2 and -8.4 kcal/mol across different versions of Moco with DMSO as substrate. 
This opposite shift of the oxidized state in relation to Moco can be directly connected to the theoretical description of the {substrate} and its reduced form (DMSO and DMS, as well as NO$_3^-$ and NO$_2^-$) with and without CPCM. 

Fig.~\ref{fig:barrier} shows the impact of the theoretical methodology employed on the energy barrier for substrate binding (E$_{\rm TS1}-$E$_{\rm Moco}$) and for substrate reduction (E$_{\rm TS2}-$E$_{\rm IM}$).
Despite the large changes in the structure of the model systems studied, the overall trend of the energy barriers with respect to the methods is qualitatively the same.
Only pCCD yields a significantly different energy barrier for DMSO binding, while all remaining CC-based methods feature essentially no energetic variation.
On the other hand, the energy barrier for binding NO$_3^-$ differs for pCCD-based and canonical CC-based methods but does not depend on the variant used.

Similarly, the method-specific trend in energy barriers for the oxidation step is qualitative comparable for both the change in the ligand and for the change in substrate.
Specifically, similar values for the barrier heights are obtained for fpLCCD, fpCCD and CCSD.
Considerably higher values are predicted by pCCD and CCD, indicating that single excitations (within a canonical basis) are required to model the transition state.
Such conclusions align with the work of Andrejić and Mata on Moco(CH$_3$O)-DMSO and Moco(PhO)-DMSO.~\cite{mata-jctc-2014}
Finally, the barrier is reduced if perturbative triple excitations are included, in particular for the reaction models containing NO$_3^-$.


\begin{table*}
\small
  \caption{
{Statistical analysis, including mean error (ME), standard deviation (SD), and mean absolute deviation (MAD), calculated using the DLPNO-CCSD(T) results as a reference for each substrate and the PR, FR, and both models combined. The definitions of ME, SD, and MAD are printed in the table footnote. The DMSO (NO$_3^-$) data set indicates error measures for the FR and PR models and DMSO (NO$_3^-$).}
}
  \label{tbl:errors}
  \begin{tabular*}{\textwidth}{@{\extracolsep{\fill}}ll  r r r r r r}
    \hline
     Data set & Error measure & pCCD  & fpLCCD & fpCCD & CCD & CCSD & DLPNO-CCSD \\
    \hline
                    &   ME  & 13.0 &	0.7	& -0.4 &	2.4	& 2.1 &	1.7     \\
\textbf{FR DMSO}    &   SD  & 6.7	&  3.7	& 6.6  &	1.5 &	1.1 &	1.0  \\
                    &   MAD & 13.0	& 3.2	& 5.2 & 	2.4	& 2.1 &	1.7 \\
    \hline   
                    &   ME  & 10.1 &	3.8 & 4.1 & 4.2 & 2.6 &	2.3    \\
\textbf{FR NO$_3^-$}&   SD  & 8.0	&   4.2	& 4.7 &	2.9	& 1.4 &	1.6  \\
                    &   MAD & 10.7 &	5.1	& 5.6 &	4.2	& 2.6 &	2.3 \\
    \hline
                    &   ME  & 11.5 &	2.2	 &  1.8	  & 3.3	& 2.3 &	2.0    \\
\textbf{FR}         &   SD  & 7.4  &	4.2	 &  6.1	  & 2.4	& 1.3 &	1.3\\
                    &   MAD & 11.8 &	4.2	 &  5.4	  & 3.3	& 2.3 &	2.0 \\
    \hline\hline
                    &   ME &  	10.0 &	0.3 &	-0.9 &	1.8 &	1.8 &	1.1 \\
\textbf{PR DMSO}    &   SD &   6.2	 &  3.4	&   6.3	 &  1.3	&   1.3	&   1.1 \\
                    &   MAD & 	10.2 &	2.7 &	4.6  &	1.9 &	1.8 &	1.2 \\
    \hline
                    &   ME &  8.1 &	 3.0 &	3.2 &	3.7	& 2.7 &	1.7 \\
\textbf{PR NO$_3^-$}&   SD&  9.0 &	3.6  &	4.1 &	2.4	& 1.3 &	1.3\\
                    &   MAD & 10.1 &	4.1 &	4.5 &	3.7	 & 2.7 &	1.7 \\
    \hline
                    &   ME  & 9.0  & 	1.7	  & 1.1   &	2.8   &  2.2  &	1.4  \\	
\textbf{PR NO$_3^-$}&   SD  & 7.5	&   3.6	  & 5.5	  & 2.1	  & 1.3	  & 1.2 \\
                    &   MAD & 10.2 &	3.4   & 4.5	  & 2.8	  & 2.2	  & 1.5	 \\
    \hline\hline
                    &   ME & 11.8	& 0.5  &	-0.6 &	2.2& 	2.0& 	1.5\\
\textbf{DMSO}       &   SD & 6.5	& 3.5  &	6.3  &	1.4 &	1.2 &	1.1\\
                    &   MAD & 11.9 &	3.0	& 4.9	& 2.2	& 2.0	& 1.5\\
    \hline
                    &   ME & 9.3 &	3.5	& 3.7 &	4.0& 	2.6&	2.1\\
\textbf{NO$_3^-$}   &   SD & 8.3 &	3.9	& 4.3 &	2.7& 	1.3 &	1.5\\
                    &   MAD & 10.5 & 4.7	& 5.2 & 4.0 &	2.6 &	2.1\\
    \hline
                    &   ME     &  	6.7  &	2.0	 & 1.5	& 3.1 &	2.3 & 1.8 \\
\textbf{All}        &   SD     &   8.3	 &  3.9	 & 5.8	& 2.3 &	1.3	& 1.3 \\
                    &   MAD   & 	11.2 &	3.9	 & 5.1	& 3.1 & 2.3	& 1.8 \\
    \hline
 \end{tabular*}
  \begin{tablenotes}\footnotesize
   \item[*] ME = $\sum_i^N \frac{E_i^{\rm method} - E_i^{\rm ref}}{N} $,
            SD = $\sqrt{\frac{\sum_i^N(E_i^{\rm ME} - \overline{E_i^{\rm ME}})^2}{N-1}}$ - standard deviation of the mean signed errors,
            MAD = $\sum_i^N \frac{|E_i^{\rm method} - E_i^{\rm ref}|}{N} $ 
\end{tablenotes}
\end{table*}

{A statistical analysis of the energy differences and barrier heights in terms of mean errors (ME), standard deviations (SD), and the mean absolute deviation (MAD) with respect to the DPLNO-CCSD(T) reference energies are presented in Table \mbox{\ref{tbl:errors}}.
We performed several statistical analyses: (i) a separate one for the FR and PR model, (ii) one for each substrate molecule for the FR and PR models, respectively, and (iii) a total statistical analysis for all reaction energies (indicated with all in the Table).
In the following, we will compare the errors of fpCC-type methods to the conventional CCSD ones as they are directly comparable, not enforcing the DLPNO approximation.
In general, the mean errors increase in the order fpCCD < fpLCCD < CCSD < CCD $\ll$ pCCD, while the standard deviation decreases in the order pCCD $\gg$ fpCCD > fpLCCD > CCD > CCSD.
On the other hand, the MAD behaves as the SD, namely gradually decreasing in the order pCCD $\gg$ fpCCD > fpLCCD > CCD > CCSD.
Thus, fpCC-type methods partly overshoot or undershoot reaction energies compared to the DLPNO-CCSD(T) reference values, resulting in a smaller ME, but larger SD and MAD.
Note, however, that the differences between two consecutive methods lie within chemical accuracy (excluding pCCD).
Hence, we can conclude that all CC methods yield qualitatively similar results for reaction energies and barrier heights.
Finally, we should mention that, despite yielding the largest error measures for all investigated CC flavors, pCCD provides smaller ME, SD, and MAD than the investigated approximate DFT exchange--correlation functionals (see Tables in the ESI).
    }

\begin{figure*}
       \centering
        \includegraphics[height=16cm]{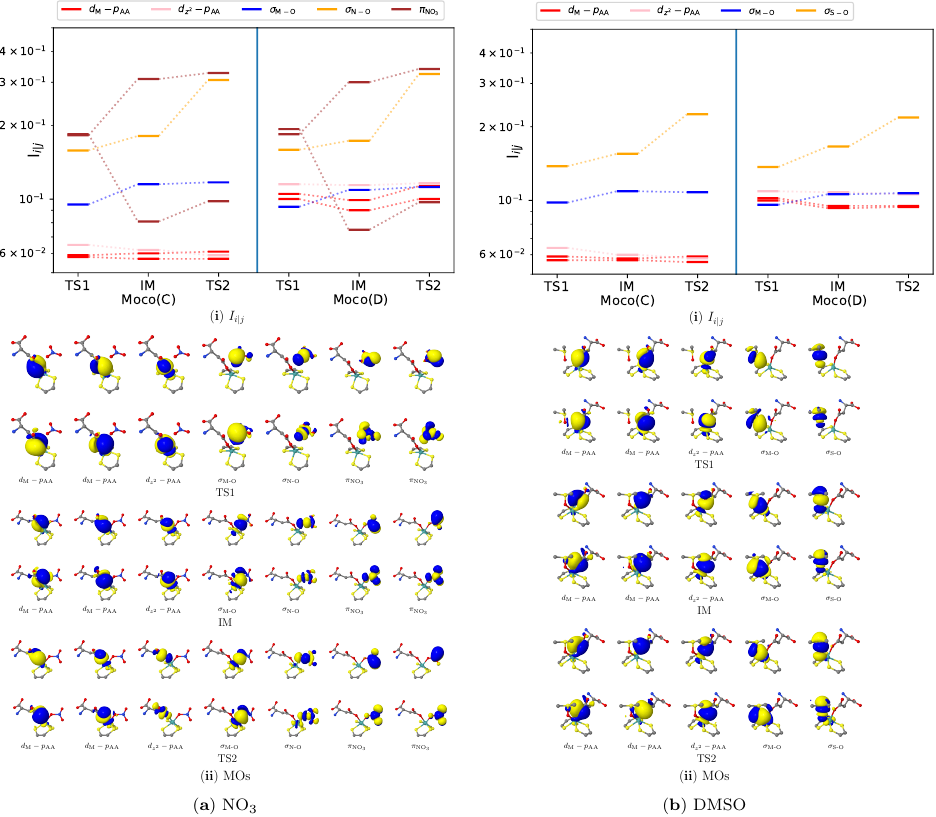}
        \caption{Orbital-pair correlations of selected orbital pairs (i) for different variants of Moco and two {substrates}, namely (a) NO$_3^-$ and (b) DMSO calculated from fpLCCD utilizing the natural pCCD-optimized orbitals (ii). Only the orbitals for the Moco(D) variant are shown in each subfigure (ii). $d_{\rm M}-p_{\rm AA}$: bonding-antibonding $d_{\rm Mo}$--$p_{{\rm O}_{\rm Asp}}$ orbital pair (AA: aminoacid), $d_{z^2}-p_{\rm AA}$: bonding-antibonding $d_{z^{2}_{\rm Mo}}$--$p_{{\rm O}_{\rm Asp}}$ orbital pair, $\sigma_{\rm M-O}$: bonding-antibonding $\sigma$ orbital pair between Mo and O$_{{\rm NO}_{3}^{-}}$/O$_{{\rm DMSO}}$, $\sigma_{\rm N-O}$/$\sigma_{\rm S-O}$: bonding-antibonding $\sigma$ orbital pair between N/S and O$_{{\rm NO}_{3}^{-}}$/O$_{{\rm DMSO}}$, $\pi_{{\rm NO}_3}$: bonding-antibonding $\pi$ orbital pair of NO$_{3}^{-}$. For a complete picture, see ESI.}
        \label{fig:correlation}
\end{figure*}

\subsection{Orbital correlation analysis}
The orbital-pair correlation analysis conducted for the Moco(C) and Moco(D) variants reacting with DMSO and NO$_3^-$ using fpLCCD and natural pCCD-optimized orbitals is depicted in Fig.~\ref{fig:correlation}. 
Our analysis focuses on selected orbitals directly involved in the bond breaking and formation processes between Moco and the {substrate}, considering the changes in orbital-pair correlations I$_{i|j}$ (between the orbital pair $i-j$) when moving along the reaction coordinate from TS1 to IM to TS2.
This includes the two bonding-antibonding $d_{\rm Mo}$--$p_{{\rm O}_{\rm AA}}$ orbital pairs (red lines), the bonding-antibonding $d_{z^{2}_{\rm Mo}}$--$p_{{\rm O}_{\rm AA}}$ orbital pair (pink lines), the bonding-antibonding $\sigma_{\rm M-O}$ orbital pair between the Mo and O$_{{\rm NO}_{3}^{-}}$/O$_{{\rm DMSO}}$ atom (blue lines), the bonding-antibonding $\sigma_{\rm N/S-O}$ orbital pair between the N/S and O$_{{\rm NO}_{3}^{-}}$/O$_{{\rm DMSO}}$ atom (orange lines), and the two bonding-antibonding $\pi$ orbital pairs of the NO$_{3}^{-}$ unit (brown lines).

For the NO$_3^-$ substrate, the most significant change in orbital-pair correlations is seen for the bonding-antibonding pair of $\pi_{{\rm NO}_3^-}$ orbitals (brown lines).
The breaking of the $\pi_{{\rm NO}_3^-}$ bonds initiates between TS1 and IM, where a simultaneous increase/decrease of the corresponding orbital-pair correlations can be observed.
This dramatic split of I$_{i|j}$ is directly linked to the rotation of the NO$_3^-$ unit, which is not symmetrically aligned with respect to the Mo center. 

Another drastic change in orbital-pair correlations can be observed for the bonding-antibonding orbital pairs involved in $\sigma_{\rm N/S-O}$ bonding (orange lines), which substantially increases from IM to TS2.
We obtain a slightly smaller orbital-pair correlation for the formation of the $\sigma_{\rm M-O}$ bond, which takes place between TS1 and IM (blue lines).
Noteworthy, the orbital-pair correlations of the Mo 4d orbitals (d$_{\rm M}$) and the p orbitals of the amino acid side chain functional group (p$_{\rm AA}$) remain rather constant along the reaction coordinate (more precisely, for the investigated critical points of the reaction path), although they are significantly more correlated in the Moco(D) variant (red and pink lines).
Finally, we observe a similar trend in I$_{i|j}$ for the DMSO substrate, albeit orbital correlations are smaller in absolute value.
To sum up, the $\sigma_{\rm M-O}$ between the substrate O atom and the Mo center is formed between TS1/IM, while the $\sigma_{\rm N/S-O}$ between the substrate O and N/S atoms of the NO${_3^-}$/DMSO unit is achieved between IM/TS2, irrespective of the substrate or Moco variant.

\section{Conclusions}\label{sec:conclusions}
In this work, we followed the assumption of Mintmier et al.\cite{mintmier_moco} that the catalytic mechanism proposed for the native Nap protein for the NO$_3^-$ reduction is followed by all residue variants (i.e., cysteine, serine, and aspartate) and substrates (NO$_3^-$ and DMSO).
Our study demonstrates that, once the substrate is bound to the Moco, the relative energies feature a low dependence on the different flavors of geometry optimization and CC-based methods.
Therefore, for a given pair of residue and substrate, the turnover energy barrier (E$_{\rm TS2}-$E$_{\rm IM}$) and unbinding barrier (E$_{\rm TS1}-$E$_{\rm IM}$) are similar for all investigated combinations of structure models and theoretical methods.
On the other hand, the energy of the free Moco compound strongly depends on the geometry optimization scheme and, to a lesser extent, to the chosen CC approximation.
As a consequence, both the binding barrier (E$_{\rm TS1}-$E$_{\rm Moco}$) and binding energy (E$_{\rm IM}-$E$_{\rm Moco}$) show similar dependencies.
This large variation in relative energies leads to a negative binding energy barrier for some combinations of geometry optimization scheme and electronic structure method.
{A statistical analysis of mean (absolute) errors and standard deviations confirms the similar performance of the investigated CC models. In general, the mean absolute deviation lies between 2-5 kcal/mol compared to the DPLNO-CCSD(T) results and increases to 11 kcal/mol for pCCD.
Based on these MAD values, the performance of CC models decreases in the order pCCD $\gg$ fpCCD > fpLCCD > CCD > CCSD > DPLNO-CCSD(T), though the discrepancies between two consecutive methods approach chemical accuracy (1 kcal/mol).
Despite its poor performance, pCCD nonetheless outperforms the investigated exchange--correlation functionals compared to the DPLNO-CCSD(T) reference (MAD(BP86) = 11 kcal/mol, MAD(B3LYP) = 14 kcal/mol, MAD(BHLYP) = 15 kcal/mol; see also ESI).}

Furthermore, we observe similar trends among all combinations of residues and substrates.
In particular, the turnover energetic barrier is similar for fpLCCD, fpCCD, CCSD, and DLPNO-CCSD, respectively.
It slightly increases for pCCD and CCD and slightly decreases for DLPNO-CCSD(T).
These results indicate that the second transition state has larger contributions of single and triple excitations than the intermediate states.
The DMSO binding energy barriers are also similar for all methods explored, except for pCCD, which over-estimates the value by more than 5~kcal/mol.
Therefore, the first transition state of Moco-DMSO has a large component of non-pair double excitations in relation to the free Moco plus DMSO.
For the NO$_3^-$ substrate, we observe an intriguing separation among methodologies.
Specifically, all pCCD-based results are similar among themselves, but different from the canonical CC-based ones, which also feature an insignificant variation with the inclusion of additional excitations.
Such a division indicates that the NO$_3^-$ binding barrier may require the relaxation of the pair-excitation amplitudes.

Finally, our results show that the turnover energy barrier for the NO$_3^-$ reduction has a substantial dependence with the ligand residue.
As such, the catalyst constant of the proposed reaction path decreases in the order cysteine, aspartate, serine---a trend that agrees with the experimental results.~\cite{mintmier_moco}
Therefore, the change in the residual should have minimal catalytic effect over the DMSO substrate reduction.
In other words, our results suggest that if the reaction path of DMSO reduction is similar to the NO$_3^-$ one, all investigated residue variants of Moco should either be active or inactive.
The similarities in the NO$_3^-$ and DMSO reaction pathways for different Moco variants are also confirmed in our orbital-pair correlation analysis of the Mo--O bond formation/O--S/N bond breaking process.
Therefore, the inclusion of the aspartate group might strongly change the geometry of the molybdenum core and/or the protein cavity, as well as the reaction path for the DMSO and, most likely, the NO$_3^-$ reduction.
Such a change could explain the activity of the Moco(D) in the presence of DMSO, but not for the TMAO, and the lower molybdenum incorporation, both observed by Mintier et al.~\cite{mintmier_moco}

Despite of our best efforts, the model system considered in this work is not capable of reproducing the experimentally observed data, either due to the lack of environmental effects or due to changes in the core and, consequently, in the model itself.
The characterization of the geometry of the Moco(D) enzyme is required to clarify both the accuracy of pCCD-based methods and the unusual activity of the aspartate mutated Moco.
More specifically, the question remains if the change in activity is mainly an electronic effect caused by the change between oxo and tiol groups to a carbonilic acid one or a geometry effect that modifies the molybdenum core, impacting its reactivity.

\section{Conflicts of interest}
There are no conflicts to declare.

\section{{Data availability}}
{The data supporting this article have been included as part of the Supplementary Information.}

\begin{acknowledgments}
M.G.~acknowledges financial support from a Ulam NAWA -- Seal of Excellence research grant (no.~BPN/SEL/2021/1/00005). 
P.T.~acknowledge financial support from the OPUS research grant from the National Science Centre, Poland (Grant No. 2019/33/B/ST4/02114). 
We acknowledge that the results of this research have been achieved using the DECI resource Bem (Grant No.~412) based in Poland at Wroclaw Centre for Networking and Supercomputing (WCSS, http://wcss.pl) with support from the PRACE aisbl. 
\includegraphics[height=0.02\textheight]{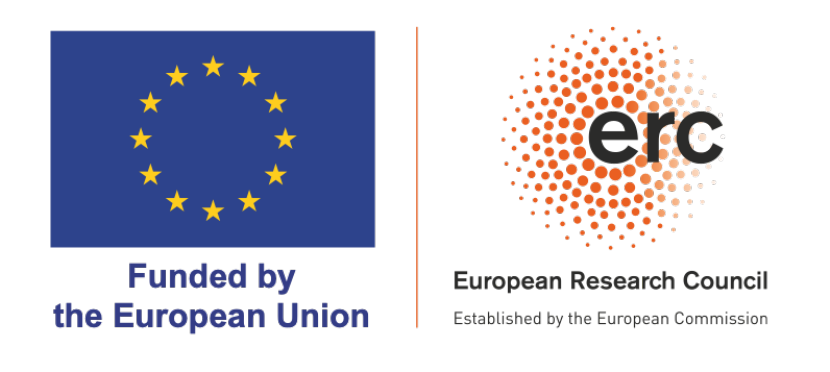} Funded/Co-funded by the European Union (ERC, DRESSED-pCCD, 101077420).
Views and opinions expressed are, however, those of the author(s) only and do not necessarily reflect those of the European Union or the European Research Council. Neither the European Union nor the granting authority can be held responsible for them. 
\end{acknowledgments}

\bibliography{rsc}

\providecommand*{\mcitethebibliography}{\thebibliography}
\csname @ifundefined\endcsname{endmcitethebibliography}
{\let\endmcitethebibliography\endthebibliography}{}
\begin{mcitethebibliography}{112}
\providecommand*{\natexlab}[1]{#1}
\providecommand*{\mciteSetBstSublistMode}[1]{}
\providecommand*{\mciteSetBstMaxWidthForm}[2]{}
\providecommand*{\mciteBstWouldAddEndPuncttrue}
  {\def\EndOfBibitem{\unskip.}}
\providecommand*{\mciteBstWouldAddEndPunctfalse}
  {\let\EndOfBibitem\relax}
\providecommand*{\mciteSetBstMidEndSepPunct}[3]{}
\providecommand*{\mciteSetBstSublistLabelBeginEnd}[3]{}
\providecommand*{\EndOfBibitem}{}
\mciteSetBstSublistMode{f}
\mciteSetBstMaxWidthForm{subitem}
{(\emph{\alph{mcitesubitemcount}})}
\mciteSetBstSublistLabelBeginEnd{\mcitemaxwidthsubitemform\space}
{\relax}{\relax}

\bibitem[Kisker \emph{et~al.}(1997)Kisker, Schindelin, and
  Rees]{kisker1997review}
C.~Kisker, H.~Schindelin and D.~C. Rees, \emph{Annu.~Rev.~Biochem.}, 1997,
  \textbf{66}, 233--267\relax
\mciteBstWouldAddEndPuncttrue
\mciteSetBstMidEndSepPunct{\mcitedefaultmidpunct}
{\mcitedefaultendpunct}{\mcitedefaultseppunct}\relax
\EndOfBibitem
\bibitem[Hille(1996)]{hill1996review}
R.~Hille, \emph{Chem.~Rev.}, 1996, \textbf{96}, 2757--2816\relax
\mciteBstWouldAddEndPuncttrue
\mciteSetBstMidEndSepPunct{\mcitedefaultmidpunct}
{\mcitedefaultendpunct}{\mcitedefaultseppunct}\relax
\EndOfBibitem
\bibitem[Mintmier \emph{et~al.}(2021)Mintmier, McGarry, Bain, and
  Basu]{mintmier_moco}
B.~Mintmier, J.~M. McGarry, D.~J. Bain and P.~Basu,
  \emph{J.~Bio.~Inorg.~Chem.}, 2021, \textbf{26}, 13--28\relax
\mciteBstWouldAddEndPuncttrue
\mciteSetBstMidEndSepPunct{\mcitedefaultmidpunct}
{\mcitedefaultendpunct}{\mcitedefaultseppunct}\relax
\EndOfBibitem
\bibitem[Johnson and Rajagopalan(2001)]{johnson2001Moco}
K.~E. Johnson and K.~V. Rajagopalan, \emph{J.~Biol.~Chem.}, 2001, \textbf{276},
  13178--13185\relax
\mciteBstWouldAddEndPuncttrue
\mciteSetBstMidEndSepPunct{\mcitedefaultmidpunct}
{\mcitedefaultendpunct}{\mcitedefaultseppunct}\relax
\EndOfBibitem
\bibitem[Moula \emph{et~al.}(2013)Moula, Bose, and Sarkar]{moula-ic-2013}
G.~Moula, M.~Bose and S.~Sarkar, \emph{Inorg.~Chem.}, 2013, \textbf{52},
  5316--5327\relax
\mciteBstWouldAddEndPuncttrue
\mciteSetBstMidEndSepPunct{\mcitedefaultmidpunct}
{\mcitedefaultendpunct}{\mcitedefaultseppunct}\relax
\EndOfBibitem
\bibitem[Lim \emph{et~al.}(2000)Lim, Sung, and Holm]{Lim2000}
B.~S. Lim, K.-M. Sung and R.~H. Holm, \emph{J.~Am.~Chem.~Soc.}, 2000,
  \textbf{122}, 7410--7411\relax
\mciteBstWouldAddEndPuncttrue
\mciteSetBstMidEndSepPunct{\mcitedefaultmidpunct}
{\mcitedefaultendpunct}{\mcitedefaultseppunct}\relax
\EndOfBibitem
\bibitem[Webster and Hall(2001)]{webster2001Moco}
C.~E. Webster and M.~B. Hall, \emph{J.~Am.~Chem.~Soc.}, 2001, \textbf{123},
  5820--5821\relax
\mciteBstWouldAddEndPuncttrue
\mciteSetBstMidEndSepPunct{\mcitedefaultmidpunct}
{\mcitedefaultendpunct}{\mcitedefaultseppunct}\relax
\EndOfBibitem
\bibitem[Thapper \emph{et~al.}(2002)Thapper, Deeth, and
  Nordlander]{thapper2002Moco}
A.~Thapper, R.~J. Deeth and E.~Nordlander, \emph{Inorg.~Chem.}, 2002,
  \textbf{41}, 6695--6702\relax
\mciteBstWouldAddEndPuncttrue
\mciteSetBstMidEndSepPunct{\mcitedefaultmidpunct}
{\mcitedefaultendpunct}{\mcitedefaultseppunct}\relax
\EndOfBibitem
\bibitem[van Severen \emph{et~al.}(2014)van Severen, Feldt, Li, Starke, Mata,
  Nordlander, and Ryde]{severen2014}
M.-C. van Severen, M.~Feldt, J.~Li, K.~Starke, R.~Mata, E.~Nordlander and
  U.~Ryde, \emph{J.~Bio.~Inorg.~Chem.}, 2014, \textbf{19}, 1165--1179\relax
\mciteBstWouldAddEndPuncttrue
\mciteSetBstMidEndSepPunct{\mcitedefaultmidpunct}
{\mcitedefaultendpunct}{\mcitedefaultseppunct}\relax
\EndOfBibitem
\bibitem[Ryde \emph{et~al.}(2009)Ryde, Schulzke, and Starke]{ryde2009}
U.~Ryde, C.~Schulzke and K.~Starke, \emph{J.~Bio.~Inorg.~Chem.}, 2009,
  \textbf{14}, 1053--1064\relax
\mciteBstWouldAddEndPuncttrue
\mciteSetBstMidEndSepPunct{\mcitedefaultmidpunct}
{\mcitedefaultendpunct}{\mcitedefaultseppunct}\relax
\EndOfBibitem
\bibitem[Hofmann(2007)]{hofmann2007}
M.~Hofmann, \emph{J.~Bio.~Inorg.~Chem.}, 2007, \textbf{12}, 989--1001\relax
\mciteBstWouldAddEndPuncttrue
\mciteSetBstMidEndSepPunct{\mcitedefaultmidpunct}
{\mcitedefaultendpunct}{\mcitedefaultseppunct}\relax
\EndOfBibitem
\bibitem[Dong and Ryde(2017)]{dong2017}
G.~Dong and U.~Ryde, \emph{J.~Inorg.~Biochem.}, 2017, \textbf{171},
  45--51\relax
\mciteBstWouldAddEndPuncttrue
\mciteSetBstMidEndSepPunct{\mcitedefaultmidpunct}
{\mcitedefaultendpunct}{\mcitedefaultseppunct}\relax
\EndOfBibitem
\bibitem[Li \emph{et~al.}(2013)Li, Mata, and Ryde]{li2013}
J.-L. Li, R.~A. Mata and U.~Ryde, \emph{J.~Chem.~Theory~Comput.}, 2013,
  \textbf{9}, 1799--1807\relax
\mciteBstWouldAddEndPuncttrue
\mciteSetBstMidEndSepPunct{\mcitedefaultmidpunct}
{\mcitedefaultendpunct}{\mcitedefaultseppunct}\relax
\EndOfBibitem
\bibitem[Li and Ryde(2014)]{li2014}
J.~Li and U.~Ryde, \emph{Inorg.~Chem.}, 2014, \textbf{53}, 11913--11924\relax
\mciteBstWouldAddEndPuncttrue
\mciteSetBstMidEndSepPunct{\mcitedefaultmidpunct}
{\mcitedefaultendpunct}{\mcitedefaultseppunct}\relax
\EndOfBibitem
\bibitem[Andrejić and Mata(2014)]{mata-jctc-2014}
M.~Andrejić and R.~A. Mata, \emph{J.~Chem.~Theory~Comput.}, 2014, \textbf{10},
  5397--5404\relax
\mciteBstWouldAddEndPuncttrue
\mciteSetBstMidEndSepPunct{\mcitedefaultmidpunct}
{\mcitedefaultendpunct}{\mcitedefaultseppunct}\relax
\EndOfBibitem
\bibitem[Tecmer and Boguslawski(2022)]{tecmer2022geminal}
P.~Tecmer and K.~Boguslawski, \emph{Phys.~Chem.~Chem.~Phys.}, 2022,
  \textbf{24}, 23026--23048\relax
\mciteBstWouldAddEndPuncttrue
\mciteSetBstMidEndSepPunct{\mcitedefaultmidpunct}
{\mcitedefaultendpunct}{\mcitedefaultseppunct}\relax
\EndOfBibitem
\bibitem[Limacher \emph{et~al.}(2013)Limacher, Ayers, Johnson, De~Baerdemacker,
  Van~Neck, and Bultinck]{limacher2013new}
P.~A. Limacher, P.~W. Ayers, P.~A. Johnson, S.~De~Baerdemacker, D.~Van~Neck and
  P.~Bultinck, \emph{J.~Chem.~Theory~Comput.}, 2013, \textbf{9},
  1394--1401\relax
\mciteBstWouldAddEndPuncttrue
\mciteSetBstMidEndSepPunct{\mcitedefaultmidpunct}
{\mcitedefaultendpunct}{\mcitedefaultseppunct}\relax
\EndOfBibitem
\bibitem[Boguslawski \emph{et~al.}(2014)Boguslawski, Tecmer, Ayers, Bultinck,
  De~Baerdemacker, and Van~Neck]{boguslawski2014efficient}
K.~Boguslawski, P.~Tecmer, P.~W. Ayers, P.~Bultinck, S.~De~Baerdemacker and
  D.~Van~Neck, \emph{Phys.~Rev.~B}, 2014, \textbf{89}, 201106\relax
\mciteBstWouldAddEndPuncttrue
\mciteSetBstMidEndSepPunct{\mcitedefaultmidpunct}
{\mcitedefaultendpunct}{\mcitedefaultseppunct}\relax
\EndOfBibitem
\bibitem[Stein \emph{et~al.}(2014)Stein, Henderson, and
  Scuseria]{stein2014seniority}
T.~Stein, T.~M. Henderson and G.~E. Scuseria, \emph{J.~Chem.~Phys.}, 2014,
  \textbf{140}, 214113\relax
\mciteBstWouldAddEndPuncttrue
\mciteSetBstMidEndSepPunct{\mcitedefaultmidpunct}
{\mcitedefaultendpunct}{\mcitedefaultseppunct}\relax
\EndOfBibitem
\bibitem[Hurley \emph{et~al.}(1953)Hurley, Lennard-Jones, and
  Pople]{hurley1953molecular}
A.~C. Hurley, J.~E. Lennard-Jones and J.~A. Pople, \emph{Proc. R. Soc. A},
  1953, \textbf{220}, 446--455\relax
\mciteBstWouldAddEndPuncttrue
\mciteSetBstMidEndSepPunct{\mcitedefaultmidpunct}
{\mcitedefaultendpunct}{\mcitedefaultseppunct}\relax
\EndOfBibitem
\bibitem[Parr \emph{et~al.}(1956)Parr, Ellison, and Lykos]{parr1956generalized}
R.~G. Parr, F.~O. Ellison and P.~G. Lykos, \emph{J.~Chem.~Phys.}, 1956,
  \textbf{24}, 1106--1106\relax
\mciteBstWouldAddEndPuncttrue
\mciteSetBstMidEndSepPunct{\mcitedefaultmidpunct}
{\mcitedefaultendpunct}{\mcitedefaultseppunct}\relax
\EndOfBibitem
\bibitem[Bardeen \emph{et~al.}(1957)Bardeen, Cooper, and
  Schrieffer]{bardeen1957theory}
J.~Bardeen, L.~N. Cooper and J.~R. Schrieffer, \emph{Phys.~Rev.}, 1957,
  \textbf{108}, 1175\relax
\mciteBstWouldAddEndPuncttrue
\mciteSetBstMidEndSepPunct{\mcitedefaultmidpunct}
{\mcitedefaultendpunct}{\mcitedefaultseppunct}\relax
\EndOfBibitem
\bibitem[Parks and Parr(1958)]{parks1958theory}
J.~M. Parks and R.~G. Parr, \emph{J.~Chem.~Phys.}, 1958, \textbf{28},
  335--345\relax
\mciteBstWouldAddEndPuncttrue
\mciteSetBstMidEndSepPunct{\mcitedefaultmidpunct}
{\mcitedefaultendpunct}{\mcitedefaultseppunct}\relax
\EndOfBibitem
\bibitem[Coleman(1965)]{coleman1965structure}
A.~Coleman, \emph{J.~Math.~Phys.}, 1965, \textbf{6}, 1425--1431\relax
\mciteBstWouldAddEndPuncttrue
\mciteSetBstMidEndSepPunct{\mcitedefaultmidpunct}
{\mcitedefaultendpunct}{\mcitedefaultseppunct}\relax
\EndOfBibitem
\bibitem[Miller and Ruedenberg(1968)]{miller1968electron}
K.~J. Miller and K.~Ruedenberg, \emph{J.~Chem.~Phys.}, 1968, \textbf{48},
  3444--3449\relax
\mciteBstWouldAddEndPuncttrue
\mciteSetBstMidEndSepPunct{\mcitedefaultmidpunct}
{\mcitedefaultendpunct}{\mcitedefaultseppunct}\relax
\EndOfBibitem
\bibitem[Surj{\'a}n(1999)]{surjan1999introduction}
P.~R. Surj{\'a}n, in \emph{Correlation and Localization}, ed. P.~R. Surj{\'a}n,
  R.~J. Bartlett, F.~Bog{\'a}r, D.~L. Cooper, B.~Kirtman, W.~Klopper,
  W.~Kutzelnigg, N.~H. March, P.~G. Mezey, H.~M{\"u}ller, J.~Noga, J.~Paldus,
  J.~Pipek, M.~Raimondi, I.~R{\o}eggen, J.~Q. Sun, P.~R. Surj{\'a}n,
  C.~Valdemoro and S.~Vogtner, Springer Berlin Heidelberg, Berlin, Heidelberg,
  1999, pp. 63--88\relax
\mciteBstWouldAddEndPuncttrue
\mciteSetBstMidEndSepPunct{\mcitedefaultmidpunct}
{\mcitedefaultendpunct}{\mcitedefaultseppunct}\relax
\EndOfBibitem
\bibitem[Surj{\'a}n \emph{et~al.}(2012)Surj{\'a}n, Szabados, Jeszenszki, and
  Zoboki]{surjan2012strongly}
P.~R. Surj{\'a}n, {\'A}.~Szabados, P.~Jeszenszki and T.~Zoboki,
  \emph{J.~Math.~Chem.}, 2012, \textbf{50}, 534--551\relax
\mciteBstWouldAddEndPuncttrue
\mciteSetBstMidEndSepPunct{\mcitedefaultmidpunct}
{\mcitedefaultendpunct}{\mcitedefaultseppunct}\relax
\EndOfBibitem
\bibitem[Tecmer \emph{et~al.}(2014)Tecmer, Boguslawski, Johnson, Limacher,
  Chan, Verstraelen, and Ayers]{tecmer2014assessing}
P.~Tecmer, K.~Boguslawski, P.~A. Johnson, P.~A. Limacher, M.~Chan,
  T.~Verstraelen and P.~W. Ayers, \emph{J.~Phys.~Chem.~A}, 2014, \textbf{118},
  9058--9068\relax
\mciteBstWouldAddEndPuncttrue
\mciteSetBstMidEndSepPunct{\mcitedefaultmidpunct}
{\mcitedefaultendpunct}{\mcitedefaultseppunct}\relax
\EndOfBibitem
\bibitem[Johnson \emph{et~al.}(2013)Johnson, Ayers, Limacher, De~Baerdemacker,
  Van~Neck, and Bultinck]{johnson2013size}
P.~A. Johnson, P.~W. Ayers, P.~A. Limacher, S.~De~Baerdemacker, D.~Van~Neck and
  P.~Bultinck, \emph{Comput.~Theory~Chem.}, 2013, \textbf{1003}, 101--113\relax
\mciteBstWouldAddEndPuncttrue
\mciteSetBstMidEndSepPunct{\mcitedefaultmidpunct}
{\mcitedefaultendpunct}{\mcitedefaultseppunct}\relax
\EndOfBibitem
\bibitem[Johnson \emph{et~al.}(2017)Johnson, Limacher, Kim, Richer,
  Miranda-Quintana, Heidar-Zadeh, Ayers, Bultinck, De~Baerdemacker, and
  Van~Neck]{johnson2017strategies}
P.~A. Johnson, P.~A. Limacher, T.~D. Kim, M.~Richer, R.~A. Miranda-Quintana,
  F.~Heidar-Zadeh, P.~W. Ayers, P.~Bultinck, S.~De~Baerdemacker and
  D.~Van~Neck, \emph{Comput.~Theory~Chem.}, 2017, \textbf{1116}, 207--219\relax
\mciteBstWouldAddEndPuncttrue
\mciteSetBstMidEndSepPunct{\mcitedefaultmidpunct}
{\mcitedefaultendpunct}{\mcitedefaultseppunct}\relax
\EndOfBibitem
\bibitem[Fecteau \emph{et~al.}(2020)Fecteau, Fortin, Cloutier, and
  Johnson]{fecteau2020reduced}
C.-{\'E}. Fecteau, H.~Fortin, S.~Cloutier and P.~A. Johnson,
  \emph{J.~Chem.~Phys.}, 2020, \textbf{153}, 164117\relax
\mciteBstWouldAddEndPuncttrue
\mciteSetBstMidEndSepPunct{\mcitedefaultmidpunct}
{\mcitedefaultendpunct}{\mcitedefaultseppunct}\relax
\EndOfBibitem
\bibitem[Johnson \emph{et~al.}(2020)Johnson, Fecteau, Berthiaume, Cloutier,
  Carrier, Gratton, Bultinck, De~Baerdemacker, Van~Neck,
  Limacher,\emph{et~al.}]{johnson2020richardson}
P.~A. Johnson, C.-{\'E}. Fecteau, F.~Berthiaume, S.~Cloutier, L.~Carrier,
  M.~Gratton, P.~Bultinck, S.~De~Baerdemacker, D.~Van~Neck, P.~Limacher
  \emph{et~al.}, \emph{J.~Chem.~Phys.}, 2020, \textbf{153}, 104110\relax
\mciteBstWouldAddEndPuncttrue
\mciteSetBstMidEndSepPunct{\mcitedefaultmidpunct}
{\mcitedefaultendpunct}{\mcitedefaultseppunct}\relax
\EndOfBibitem
\bibitem[Johnson \emph{et~al.}(2022)Johnson, Ayers, De~Baerdemacker, Limacher,
  and Van~Neck]{johnson2022bivariational}
P.~A. Johnson, P.~W. Ayers, S.~De~Baerdemacker, P.~A. Limacher and D.~Van~Neck,
  \emph{Comput.~Theory~Chem.}, 2022, \textbf{1212}, 113718\relax
\mciteBstWouldAddEndPuncttrue
\mciteSetBstMidEndSepPunct{\mcitedefaultmidpunct}
{\mcitedefaultendpunct}{\mcitedefaultseppunct}\relax
\EndOfBibitem
\bibitem[Faribault \emph{et~al.}(2022)Faribault, Dimo, Moisset, and
  Johnson]{faribault2022reduced}
A.~Faribault, C.~Dimo, J.-D. Moisset and P.~A. Johnson, \emph{J.~Chem.~Phys.},
  2022, \textbf{157}, 214104\relax
\mciteBstWouldAddEndPuncttrue
\mciteSetBstMidEndSepPunct{\mcitedefaultmidpunct}
{\mcitedefaultendpunct}{\mcitedefaultseppunct}\relax
\EndOfBibitem
\bibitem[Fecteau \emph{et~al.}(2022)Fecteau, Cloutier, Moisset, Boulay,
  Bultinck, Faribault, and Johnson]{fecteau2022near}
C.-{\'E}. Fecteau, S.~Cloutier, J.-D. Moisset, J.~Boulay, P.~Bultinck,
  A.~Faribault and P.~A. Johnson, \emph{J.~Chem.~Phys.}, 2022, \textbf{156},
  194103\relax
\mciteBstWouldAddEndPuncttrue
\mciteSetBstMidEndSepPunct{\mcitedefaultmidpunct}
{\mcitedefaultendpunct}{\mcitedefaultseppunct}\relax
\EndOfBibitem
\bibitem[Moisset \emph{et~al.}(2022)Moisset, Fecteau, and
  Johnson]{moisset2022density}
J.-D. Moisset, C.-{\'E}. Fecteau and P.~A. Johnson, \emph{J.~Chem.~Phys.},
  2022, \textbf{156}, 214110\relax
\mciteBstWouldAddEndPuncttrue
\mciteSetBstMidEndSepPunct{\mcitedefaultmidpunct}
{\mcitedefaultendpunct}{\mcitedefaultseppunct}\relax
\EndOfBibitem
\bibitem[Limacher \emph{et~al.}(2014)Limacher, Kim, Ayers, Johnson,
  De~Baerdemacker, Van~Neck, and Bultinck]{limacher2014influence}
P.~A. Limacher, T.~D. Kim, P.~W. Ayers, P.~A. Johnson, S.~De~Baerdemacker,
  D.~Van~Neck and P.~Bultinck, \emph{Mol.~Phys.}, 2014, \textbf{112},
  853--862\relax
\mciteBstWouldAddEndPuncttrue
\mciteSetBstMidEndSepPunct{\mcitedefaultmidpunct}
{\mcitedefaultendpunct}{\mcitedefaultseppunct}\relax
\EndOfBibitem
\bibitem[Boguslawski \emph{et~al.}(2014)Boguslawski, Tecmer, Bultinck,
  De~Baerdemacker, Van~Neck, and Ayers]{boguslawski2014nonvariational}
K.~Boguslawski, P.~Tecmer, P.~Bultinck, S.~De~Baerdemacker, D.~Van~Neck and
  P.~W. Ayers, \emph{J.~Chem.~Theory~Comput.}, 2014, \textbf{10},
  4873--4882\relax
\mciteBstWouldAddEndPuncttrue
\mciteSetBstMidEndSepPunct{\mcitedefaultmidpunct}
{\mcitedefaultendpunct}{\mcitedefaultseppunct}\relax
\EndOfBibitem
\bibitem[Boguslawski \emph{et~al.}(2014)Boguslawski, Tecmer, Limacher, Johnson,
  Ayers, Bultinck, De~Baerdemacker, and Van~Neck]{boguslawski2014projected}
K.~Boguslawski, P.~Tecmer, P.~A. Limacher, P.~A. Johnson, P.~W. Ayers,
  P.~Bultinck, S.~De~Baerdemacker and D.~Van~Neck, \emph{J.~Chem.~Phys.}, 2014,
  \textbf{140}, 214114\relax
\mciteBstWouldAddEndPuncttrue
\mciteSetBstMidEndSepPunct{\mcitedefaultmidpunct}
{\mcitedefaultendpunct}{\mcitedefaultseppunct}\relax
\EndOfBibitem
\bibitem[Boguslawski \emph{et~al.}(2016)Boguslawski, Tecmer, and
  Legeza]{boguslawski2016analysis}
K.~Boguslawski, P.~Tecmer and {\"O}.~Legeza, \emph{Phys.~Rev.~B}, 2016,
  \textbf{94}, 155126\relax
\mciteBstWouldAddEndPuncttrue
\mciteSetBstMidEndSepPunct{\mcitedefaultmidpunct}
{\mcitedefaultendpunct}{\mcitedefaultseppunct}\relax
\EndOfBibitem
\bibitem[Limacher(2015)]{limacher2015orbital}
P.~A. Limacher, \emph{J.~Chem.~Theory~Comput.}, 2015, \textbf{11},
  3629--3635\relax
\mciteBstWouldAddEndPuncttrue
\mciteSetBstMidEndSepPunct{\mcitedefaultmidpunct}
{\mcitedefaultendpunct}{\mcitedefaultseppunct}\relax
\EndOfBibitem
\bibitem[Tecmer \emph{et~al.}(2015)Tecmer, Boguslawski, and
  Ayers]{tecmer2015singlet}
P.~Tecmer, K.~Boguslawski and P.~W. Ayers, \emph{Phys.~Chem.~Chem.~Phys.},
  2015, \textbf{17}, 14427--14436\relax
\mciteBstWouldAddEndPuncttrue
\mciteSetBstMidEndSepPunct{\mcitedefaultmidpunct}
{\mcitedefaultendpunct}{\mcitedefaultseppunct}\relax
\EndOfBibitem
\bibitem[Brz{\k{e}}k \emph{et~al.}(2019)Brz{\k{e}}k, Boguslawski, Tecmer, and
  {\.Z}uchowski]{brzek2019benchmarking}
F.~Brz{\k{e}}k, K.~Boguslawski, P.~Tecmer and P.~S. {\.Z}uchowski,
  \emph{J.~Chem.~Theory~Comput.}, 2019, \textbf{15}, 4021--4035\relax
\mciteBstWouldAddEndPuncttrue
\mciteSetBstMidEndSepPunct{\mcitedefaultmidpunct}
{\mcitedefaultendpunct}{\mcitedefaultseppunct}\relax
\EndOfBibitem
\bibitem[Henderson and Scuseria(2019)]{henderson2019geminal}
T.~M. Henderson and G.~E. Scuseria, \emph{J.~Chem.~Phys.}, 2019, \textbf{151},
  051101\relax
\mciteBstWouldAddEndPuncttrue
\mciteSetBstMidEndSepPunct{\mcitedefaultmidpunct}
{\mcitedefaultendpunct}{\mcitedefaultseppunct}\relax
\EndOfBibitem
\bibitem[Nowak \emph{et~al.}(2021)Nowak, Legeza, and
  Boguslawski]{nowak2021orbital}
A.~Nowak, {\"O}.~Legeza and K.~Boguslawski, \emph{J.~Chem.~Phys.}, 2021,
  \textbf{154}, 084111\relax
\mciteBstWouldAddEndPuncttrue
\mciteSetBstMidEndSepPunct{\mcitedefaultmidpunct}
{\mcitedefaultendpunct}{\mcitedefaultseppunct}\relax
\EndOfBibitem
\bibitem[Leszczyk \emph{et~al.}(2021)Leszczyk, M{\'a}t{\'e}, Legeza, and
  Boguslawski]{leszczyk2021assessing}
A.~Leszczyk, M.~M{\'a}t{\'e}, O.~Legeza and K.~Boguslawski,
  \emph{J.~Chem.~Theory~Comput.}, 2021, \textbf{18}, 96--117\relax
\mciteBstWouldAddEndPuncttrue
\mciteSetBstMidEndSepPunct{\mcitedefaultmidpunct}
{\mcitedefaultendpunct}{\mcitedefaultseppunct}\relax
\EndOfBibitem
\bibitem[Leszczyk \emph{et~al.}(2022)Leszczyk, Dome, Tecmer, Kedziera, and
  Boguslawski]{leszczyk2022}
A.~Leszczyk, T.~Dome, P.~Tecmer, D.~Kedziera and K.~Boguslawski, \emph{Phys.
  Chem. Chem. Phys.}, 2022, \textbf{24}, 21296--21307\relax
\mciteBstWouldAddEndPuncttrue
\mciteSetBstMidEndSepPunct{\mcitedefaultmidpunct}
{\mcitedefaultendpunct}{\mcitedefaultseppunct}\relax
\EndOfBibitem
\bibitem[Chakraborty \emph{et~al.}(2024)Chakraborty, de~Moraes, Boguslawski,
  Nowak, Świerczynski, and Tecmer]{pccd-dipole-moments-2024}
R.~Chakraborty, M.~M.~F. de~Moraes, K.~Boguslawski, A.~Nowak, J.~Świerczynski
  and P.~Tecmer, \emph{J.~Chem.~Theory~Comput.}, 2024, \textbf{20},
  4689–4702\relax
\mciteBstWouldAddEndPuncttrue
\mciteSetBstMidEndSepPunct{\mcitedefaultmidpunct}
{\mcitedefaultendpunct}{\mcitedefaultseppunct}\relax
\EndOfBibitem
\bibitem[Tecmer \emph{et~al.}(2019)Tecmer, Boguslawski, Borkowski,
  {\.Z}uchowski, and K{\k{e}}dziera]{tecmer2019modeling}
P.~Tecmer, K.~Boguslawski, M.~Borkowski, P.~S. {\.Z}uchowski and
  D.~K{\k{e}}dziera, \emph{Int.~J.~Quantum~Chem.}, 2019, \textbf{119},
  e25983\relax
\mciteBstWouldAddEndPuncttrue
\mciteSetBstMidEndSepPunct{\mcitedefaultmidpunct}
{\mcitedefaultendpunct}{\mcitedefaultseppunct}\relax
\EndOfBibitem
\bibitem[Garza \emph{et~al.}(2015)Garza, Sousa~Alencar, and
  Scuseria]{garza2015actinide}
A.~J. Garza, A.~G. Sousa~Alencar and G.~E. Scuseria, \emph{J.~Chem.~Phys.},
  2015, \textbf{143}, 244106\relax
\mciteBstWouldAddEndPuncttrue
\mciteSetBstMidEndSepPunct{\mcitedefaultmidpunct}
{\mcitedefaultendpunct}{\mcitedefaultseppunct}\relax
\EndOfBibitem
\bibitem[Boguslawski(2016)]{boguslawski2016targeting}
K.~Boguslawski, \emph{J.~Chem.~Phys.}, 2016, \textbf{145}, 234105\relax
\mciteBstWouldAddEndPuncttrue
\mciteSetBstMidEndSepPunct{\mcitedefaultmidpunct}
{\mcitedefaultendpunct}{\mcitedefaultseppunct}\relax
\EndOfBibitem
\bibitem[Boguslawski(2017)]{boguslawski2017erratum}
K.~Boguslawski, \emph{J.~Chem.~Phys.}, 2017, \textbf{147}, 139901\relax
\mciteBstWouldAddEndPuncttrue
\mciteSetBstMidEndSepPunct{\mcitedefaultmidpunct}
{\mcitedefaultendpunct}{\mcitedefaultseppunct}\relax
\EndOfBibitem
\bibitem[Nowak \emph{et~al.}(2019)Nowak, Tecmer, and
  Boguslawski]{nowak2019assessing}
A.~Nowak, P.~Tecmer and K.~Boguslawski, \emph{Phys.~Chem.~Chem.~Phys.}, 2019,
  \textbf{21}, 19039--19053\relax
\mciteBstWouldAddEndPuncttrue
\mciteSetBstMidEndSepPunct{\mcitedefaultmidpunct}
{\mcitedefaultendpunct}{\mcitedefaultseppunct}\relax
\EndOfBibitem
\bibitem[Nowak and Boguslawski(2023)]{Nowak2023}
A.~Nowak and K.~Boguslawski, \emph{Phys. Chem. Chem. Phys.}, 2023, \textbf{25},
  7289--7301\relax
\mciteBstWouldAddEndPuncttrue
\mciteSetBstMidEndSepPunct{\mcitedefaultmidpunct}
{\mcitedefaultendpunct}{\mcitedefaultseppunct}\relax
\EndOfBibitem
\bibitem[Chakraborty \emph{et~al.}(2023)Chakraborty, Boguslawski, and
  Tecmer]{chakraborty2023}
R.~Chakraborty, K.~Boguslawski and P.~Tecmer, \emph{Phys.~Chem.~Chem.~Phys.},
  2023, \textbf{25}, 25377--25388\relax
\mciteBstWouldAddEndPuncttrue
\mciteSetBstMidEndSepPunct{\mcitedefaultmidpunct}
{\mcitedefaultendpunct}{\mcitedefaultseppunct}\relax
\EndOfBibitem
\bibitem[Jahani \emph{et~al.}(2023)Jahani, Boguslawski, and
  Tecmer]{jahani2023relationship}
S.~Jahani, K.~Boguslawski and P.~Tecmer, \emph{RSC~Adv.}, 2023, \textbf{13},
  27898--27911\relax
\mciteBstWouldAddEndPuncttrue
\mciteSetBstMidEndSepPunct{\mcitedefaultmidpunct}
{\mcitedefaultendpunct}{\mcitedefaultseppunct}\relax
\EndOfBibitem
\bibitem[Tecmer \emph{et~al.}(2023)Tecmer, Ga{\l}y{\'n}ska, Szczuczko, and
  Boguslawski]{tecmer2023jpcl}
P.~Tecmer, M.~Ga{\l}y{\'n}ska, L.~Szczuczko and K.~Boguslawski,
  \emph{J.~Phys.~Chem.~Lett.}, 2023, \textbf{14}, 9909--9917\relax
\mciteBstWouldAddEndPuncttrue
\mciteSetBstMidEndSepPunct{\mcitedefaultmidpunct}
{\mcitedefaultendpunct}{\mcitedefaultseppunct}\relax
\EndOfBibitem
\bibitem[Boguslawski(2018)]{boguslawski2018targeting}
K.~Boguslawski, \emph{J.~Chem.~Theory~Comput.}, 2018, \textbf{15}, 18--24\relax
\mciteBstWouldAddEndPuncttrue
\mciteSetBstMidEndSepPunct{\mcitedefaultmidpunct}
{\mcitedefaultendpunct}{\mcitedefaultseppunct}\relax
\EndOfBibitem
\bibitem[Kossoski \emph{et~al.}(2021)Kossoski, Marie, Scemama, Caffarel, and
  Loos]{kossoski2021excited}
F.~Kossoski, A.~Marie, A.~Scemama, M.~Caffarel and P.-F. Loos,
  \emph{J.~Chem.~Theory~Comput.}, 2021, \textbf{17}, 4756--4768\relax
\mciteBstWouldAddEndPuncttrue
\mciteSetBstMidEndSepPunct{\mcitedefaultmidpunct}
{\mcitedefaultendpunct}{\mcitedefaultseppunct}\relax
\EndOfBibitem
\bibitem[Ravi \emph{et~al.}(2023)Ravi, Perera, Park, and
  Bartlett]{bartlett-pccd-tcc}
M.~Ravi, A.~Perera, Y.~C. Park and R.~J. Bartlett, \emph{J.~Chem.~Phys.}, 2023,
  \textbf{159}, 094101\relax
\mciteBstWouldAddEndPuncttrue
\mciteSetBstMidEndSepPunct{\mcitedefaultmidpunct}
{\mcitedefaultendpunct}{\mcitedefaultseppunct}\relax
\EndOfBibitem
\bibitem[Boguslawski(2021)]{boguslawski2021open}
K.~Boguslawski, \emph{Chem.~Commun.}, 2021, \textbf{57}, 12277--12280\relax
\mciteBstWouldAddEndPuncttrue
\mciteSetBstMidEndSepPunct{\mcitedefaultmidpunct}
{\mcitedefaultendpunct}{\mcitedefaultseppunct}\relax
\EndOfBibitem
\bibitem[Mamache \emph{et~al.}(2023)Mamache, Ga{\l}y{\'n}ska, and
  Boguslawski]{mamache2023}
S.~Mamache, M.~Ga{\l}y{\'n}ska and K.~Boguslawski,
  \emph{Phys.~Chem.~Chem.~Phys.}, 2023, \textbf{25}, 18023--18029\relax
\mciteBstWouldAddEndPuncttrue
\mciteSetBstMidEndSepPunct{\mcitedefaultmidpunct}
{\mcitedefaultendpunct}{\mcitedefaultseppunct}\relax
\EndOfBibitem
\bibitem[Gałyńska and Boguslawski(2024)]{tailored-ip-pccd-jctc-2024}
M.~Gałyńska and K.~Boguslawski, \emph{J.~Chem.~Theory~Comput.}, 2024,
  \textbf{20}, 4182--4195\relax
\mciteBstWouldAddEndPuncttrue
\mciteSetBstMidEndSepPunct{\mcitedefaultmidpunct}
{\mcitedefaultendpunct}{\mcitedefaultseppunct}\relax
\EndOfBibitem
\bibitem[Sinano\u{g}lu and Tuan(1963)]{sinanoglu1963}
O.~Sinano\u{g}lu and D.~F. Tuan, \emph{J. Chem. Phys.}, 1963, \textbf{38},
  1740--1748\relax
\mciteBstWouldAddEndPuncttrue
\mciteSetBstMidEndSepPunct{\mcitedefaultmidpunct}
{\mcitedefaultendpunct}{\mcitedefaultseppunct}\relax
\EndOfBibitem
\bibitem[Bartlett and Stanton(1994)]{bartlett_1994}
R.~J. Bartlett and J.~F. Stanton, \emph{Rev. Comput. Chem.}, 1994, \textbf{5},
  165--169\relax
\mciteBstWouldAddEndPuncttrue
\mciteSetBstMidEndSepPunct{\mcitedefaultmidpunct}
{\mcitedefaultendpunct}{\mcitedefaultseppunct}\relax
\EndOfBibitem
\bibitem[Boguslawski \emph{et~al.}(2012)Boguslawski, Tecmer, Legeza, and
  Reiher]{entanglement_letter}
K.~Boguslawski, P.~Tecmer, O.~Legeza and M.~Reiher, \emph{J. Phys. Chem.
  Lett.}, 2012, \textbf{3}, 3129--3135\relax
\mciteBstWouldAddEndPuncttrue
\mciteSetBstMidEndSepPunct{\mcitedefaultmidpunct}
{\mcitedefaultendpunct}{\mcitedefaultseppunct}\relax
\EndOfBibitem
\bibitem[Henderson \emph{et~al.}(2014)Henderson, Bulik, Stein, and
  Scuseria]{henderson2014seniority}
T.~M. Henderson, I.~W. Bulik, T.~Stein and G.~E. Scuseria,
  \emph{J.~Chem.~Phys.}, 2014, \textbf{141}, 244104\relax
\mciteBstWouldAddEndPuncttrue
\mciteSetBstMidEndSepPunct{\mcitedefaultmidpunct}
{\mcitedefaultendpunct}{\mcitedefaultseppunct}\relax
\EndOfBibitem
\bibitem[Boguslawski and Ayers(2015)]{boguslawski2015linearized}
K.~Boguslawski and P.~W. Ayers, \emph{J.~Chem.~Theory~Comput.}, 2015,
  \textbf{11}, 5252--5261\relax
\mciteBstWouldAddEndPuncttrue
\mciteSetBstMidEndSepPunct{\mcitedefaultmidpunct}
{\mcitedefaultendpunct}{\mcitedefaultseppunct}\relax
\EndOfBibitem
\bibitem[Kinoshita \emph{et~al.}(2005)Kinoshita, Hino, and
  Bartlett]{kinoshita2005}
T.~Kinoshita, O.~Hino and R.~J. Bartlett, \emph{J.~Chem.~Phys.}, 2005,
  \textbf{123}, 074106\relax
\mciteBstWouldAddEndPuncttrue
\mciteSetBstMidEndSepPunct{\mcitedefaultmidpunct}
{\mcitedefaultendpunct}{\mcitedefaultseppunct}\relax
\EndOfBibitem
\bibitem[Hino \emph{et~al.}(2006)Hino, Kinoshita, Chan, and
  Bartlett]{tailoredcc2006}
O.~Hino, T.~Kinoshita, G.~K.-L. Chan and R.~J. Bartlett, \emph{J.~Chem.~Phys.},
  2006, \textbf{124}, 114311\relax
\mciteBstWouldAddEndPuncttrue
\mciteSetBstMidEndSepPunct{\mcitedefaultmidpunct}
{\mcitedefaultendpunct}{\mcitedefaultseppunct}\relax
\EndOfBibitem
\bibitem[Lyakh \emph{et~al.}(2011)Lyakh, Lotrich, and Bartlett]{lyakh2011}
D.~I. Lyakh, V.~F. Lotrich and R.~J. Bartlett, \emph{Chem.~Phys.~Lett.}, 2011,
  \textbf{501}, 166--171\relax
\mciteBstWouldAddEndPuncttrue
\mciteSetBstMidEndSepPunct{\mcitedefaultmidpunct}
{\mcitedefaultendpunct}{\mcitedefaultseppunct}\relax
\EndOfBibitem
\bibitem[Melnichuk and Bartlett(2012)]{tailoredcc2012}
A.~Melnichuk and R.~J. Bartlett, \emph{J.~Chem.~Phys.}, 2012, \textbf{137},
  214103\relax
\mciteBstWouldAddEndPuncttrue
\mciteSetBstMidEndSepPunct{\mcitedefaultmidpunct}
{\mcitedefaultendpunct}{\mcitedefaultseppunct}\relax
\EndOfBibitem
\bibitem[Boguslawski and Tecmer(2017)]{boguslawski2017benchmark}
K.~Boguslawski and P.~Tecmer, \emph{J.~Chem.~Theory~Comput.}, 2017,
  \textbf{13}, 5966--5983\relax
\mciteBstWouldAddEndPuncttrue
\mciteSetBstMidEndSepPunct{\mcitedefaultmidpunct}
{\mcitedefaultendpunct}{\mcitedefaultseppunct}\relax
\EndOfBibitem
\bibitem[Rissler \emph{et~al.}(2006)Rissler, Noack, and White]{rissler2006}
J.~Rissler, R.~M. Noack and S.~R. White, \emph{Chem. Phys.}, 2006,
  \textbf{323}, 519--531\relax
\mciteBstWouldAddEndPuncttrue
\mciteSetBstMidEndSepPunct{\mcitedefaultmidpunct}
{\mcitedefaultendpunct}{\mcitedefaultseppunct}\relax
\EndOfBibitem
\bibitem[Barcza \emph{et~al.}(2014)Barcza, Noack, S{\'o}lyom, and
  Legeza]{barcza2014entanglement}
G.~Barcza, R.~Noack, J.~S{\'o}lyom and {\"O}.~Legeza, \emph{Phys.~Rev.~B},
  2014, \textbf{92}, 125140\relax
\mciteBstWouldAddEndPuncttrue
\mciteSetBstMidEndSepPunct{\mcitedefaultmidpunct}
{\mcitedefaultendpunct}{\mcitedefaultseppunct}\relax
\EndOfBibitem
\bibitem[{K. Boguslawski, P. Tecmer}(2015)]{ijqc-2015}
{K. Boguslawski, P. Tecmer}, \emph{Int. J. Quantum Chem.}, 2015, \textbf{115},
  1289--1295\relax
\mciteBstWouldAddEndPuncttrue
\mciteSetBstMidEndSepPunct{\mcitedefaultmidpunct}
{\mcitedefaultendpunct}{\mcitedefaultseppunct}\relax
\EndOfBibitem
\bibitem[{K. Boguslawski, P. Tecmer}(2017)]{ijqc-erratum}
{K. Boguslawski, P. Tecmer}, \emph{Int. J. Quantum Chem.}, 2017, \textbf{117},
  e25455\relax
\mciteBstWouldAddEndPuncttrue
\mciteSetBstMidEndSepPunct{\mcitedefaultmidpunct}
{\mcitedefaultendpunct}{\mcitedefaultseppunct}\relax
\EndOfBibitem
\bibitem[Ding and Schilling(2020)]{Ding2020}
L.~Ding and C.~Schilling, \emph{J.~Chem.~Theory~Comput.}, 2020, \textbf{16},
  4159--4175\relax
\mciteBstWouldAddEndPuncttrue
\mciteSetBstMidEndSepPunct{\mcitedefaultmidpunct}
{\mcitedefaultendpunct}{\mcitedefaultseppunct}\relax
\EndOfBibitem
\bibitem[Ding \emph{et~al.}(2021)Ding, Mardazad, Das, Szalay, Schollw\"ock,
  Zimbor{\'a}s, and Schilling]{qit-concepts-schilling-jctc-2021}
L.~Ding, S.~Mardazad, S.~Das, S.~Szalay, U.~Schollw\"ock, Z.~Zimbor{\'a}s and
  C.~Schilling, \emph{J.~Chem.~Theory~Comput.}, 2021, \textbf{17}, 79--95\relax
\mciteBstWouldAddEndPuncttrue
\mciteSetBstMidEndSepPunct{\mcitedefaultmidpunct}
{\mcitedefaultendpunct}{\mcitedefaultseppunct}\relax
\EndOfBibitem
\bibitem[Szalay \emph{et~al.}(2015)Szalay, Pfeffer, Murg, Barcza, Verstraete,
  Schneider, and Legeza]{ors_ijqc}
S.~Szalay, M.~Pfeffer, V.~Murg, G.~Barcza, F.~Verstraete, R.~Schneider and
  {\"O}.~Legeza, \emph{Int.~J.~Quantum~Chem.}, 2015, \textbf{115},
  1342--1391\relax
\mciteBstWouldAddEndPuncttrue
\mciteSetBstMidEndSepPunct{\mcitedefaultmidpunct}
{\mcitedefaultendpunct}{\mcitedefaultseppunct}\relax
\EndOfBibitem
\bibitem[Barcza \emph{et~al.}(2011)Barcza, Legeza, Marti, and
  Reiher]{barcza_11}
G.~Barcza, O.~Legeza, K.~H. Marti and M.~Reiher, \emph{Phys. Rev. A}, 2011,
  \textbf{83}, 012508\relax
\mciteBstWouldAddEndPuncttrue
\mciteSetBstMidEndSepPunct{\mcitedefaultmidpunct}
{\mcitedefaultendpunct}{\mcitedefaultseppunct}\relax
\EndOfBibitem
\bibitem[Boguslawski \emph{et~al.}(2013)Boguslawski, Tecmer, Barcza, Legeza,
  and Reiher]{entanglement_bonding_2013}
K.~Boguslawski, P.~Tecmer, G.~Barcza, O.~Legeza and M.~Reiher, \emph{J. Chem.
  Theory Comput.}, 2013, \textbf{9}, 2959--2973\relax
\mciteBstWouldAddEndPuncttrue
\mciteSetBstMidEndSepPunct{\mcitedefaultmidpunct}
{\mcitedefaultendpunct}{\mcitedefaultseppunct}\relax
\EndOfBibitem
\bibitem[Ding \emph{et~al.}(2022)Ding, Knecht, Zimbor{\'a}s, and
  Schilling]{ding2022quantum}
L.~Ding, S.~Knecht, Z.~Zimbor{\'a}s and C.~Schilling,
  \emph{Quantum~Sci.~Technol.}, 2022, \textbf{8}, 015015\relax
\mciteBstWouldAddEndPuncttrue
\mciteSetBstMidEndSepPunct{\mcitedefaultmidpunct}
{\mcitedefaultendpunct}{\mcitedefaultseppunct}\relax
\EndOfBibitem
\bibitem[Mottet \emph{et~al.}(2014)Mottet, Tecmer, Boguslawski, Legeza, and
  Reiher]{mottet}
M.~Mottet, P.~Tecmer, K.~Boguslawski, {\"O}.~Legeza and M.~Reiher, \emph{Phys.
  Chem. Chem. Phys.}, 2014, \textbf{16}, 8872--8880\relax
\mciteBstWouldAddEndPuncttrue
\mciteSetBstMidEndSepPunct{\mcitedefaultmidpunct}
{\mcitedefaultendpunct}{\mcitedefaultseppunct}\relax
\EndOfBibitem
\bibitem[Duperrouzel \emph{et~al.}(2015)Duperrouzel, Tecmer, Boguslawski,
  Barcza, Legeza, and Ayers]{corinne_2015}
C.~Duperrouzel, P.~Tecmer, K.~Boguslawski, G.~Barcza, O.~Legeza and P.~W.
  Ayers, \emph{Chem. Phys. Lett.}, 2015, \textbf{621}, 160--164\relax
\mciteBstWouldAddEndPuncttrue
\mciteSetBstMidEndSepPunct{\mcitedefaultmidpunct}
{\mcitedefaultendpunct}{\mcitedefaultseppunct}\relax
\EndOfBibitem
\bibitem[Zhao \emph{et~al.}(2015)Zhao, Boguslawski, Tecmer, Duperrouzel,
  Barcza, Legeza, and Ayers]{zhao2015}
Y.~Zhao, K.~Boguslawski, P.~Tecmer, C.~Duperrouzel, G.~Barcza, {\"{O}}.~Legeza
  and P.~W. Ayers, \emph{Theor. Chem. Acc.}, 2015, \textbf{134}, 120\relax
\mciteBstWouldAddEndPuncttrue
\mciteSetBstMidEndSepPunct{\mcitedefaultmidpunct}
{\mcitedefaultendpunct}{\mcitedefaultseppunct}\relax
\EndOfBibitem
\bibitem[Freitag \emph{et~al.}(2015)Freitag, Knecht, Keller, Delcey, Aquilante,
  Pedersen, Lindh, Reiher, and Gonzalez]{roland-runo}
L.~Freitag, S.~Knecht, S.~F. Keller, M.~G. Delcey, F.~Aquilante, T.~B.
  Pedersen, R.~Lindh, M.~Reiher and L.~Gonzalez, \emph{Phys. Chem. Chem.
  Phys.}, 2015, \textbf{17}, 13769--13769\relax
\mciteBstWouldAddEndPuncttrue
\mciteSetBstMidEndSepPunct{\mcitedefaultmidpunct}
{\mcitedefaultendpunct}{\mcitedefaultseppunct}\relax
\EndOfBibitem
\bibitem[Stein and Reiher(2016)]{stein2016}
C.~J. Stein and M.~Reiher, \emph{J.~Chem.~Theory~Comput.}, 2016, \textbf{12},
  1760--1771\relax
\mciteBstWouldAddEndPuncttrue
\mciteSetBstMidEndSepPunct{\mcitedefaultmidpunct}
{\mcitedefaultendpunct}{\mcitedefaultseppunct}\relax
\EndOfBibitem
\bibitem[{K. Boguslawski, F. R\'eal, P. Tecmer, C. Duperrouzel, A. S. P. Gomes,
  \"{O}. Legeza, P. W. Ayers, V. Vallet}(2017)]{boguslawski2017}
{K. Boguslawski, F. R\'eal, P. Tecmer, C. Duperrouzel, A. S. P. Gomes, \"{O}.
  Legeza, P. W. Ayers, V. Vallet}, \emph{Phys. Chem. Chem. Phys.}, 2017,
  \textbf{19}, 4317--4329\relax
\mciteBstWouldAddEndPuncttrue
\mciteSetBstMidEndSepPunct{\mcitedefaultmidpunct}
{\mcitedefaultendpunct}{\mcitedefaultseppunct}\relax
\EndOfBibitem
\bibitem[Perdew(1986)]{perdew86}
J.~Perdew, \emph{Phys. Rev. B}, 1986, \textbf{33}, 8822--8824\relax
\mciteBstWouldAddEndPuncttrue
\mciteSetBstMidEndSepPunct{\mcitedefaultmidpunct}
{\mcitedefaultendpunct}{\mcitedefaultseppunct}\relax
\EndOfBibitem
\bibitem[Becke(1988)]{becke88}
A.~Becke, \emph{Phys. Rev. A}, 1988, \textbf{38}, 3098--4000\relax
\mciteBstWouldAddEndPuncttrue
\mciteSetBstMidEndSepPunct{\mcitedefaultmidpunct}
{\mcitedefaultendpunct}{\mcitedefaultseppunct}\relax
\EndOfBibitem
\bibitem[{Canal Neto} \emph{et~al.}(2021){Canal Neto}, Ferreira, Jorge, and {de
  Oliveira}]{basis_set_zora_jorge}
A.~{Canal Neto}, I.~Ferreira, F.~Jorge and A.~{de Oliveira},
  \emph{Chem.~Phys.~Lett.}, 2021, \textbf{771}, 138548\relax
\mciteBstWouldAddEndPuncttrue
\mciteSetBstMidEndSepPunct{\mcitedefaultmidpunct}
{\mcitedefaultendpunct}{\mcitedefaultseppunct}\relax
\EndOfBibitem
\bibitem[Weigend and Ahlrichs(2005)]{def2-tvzp}
F.~Weigend and R.~Ahlrichs, \emph{Phys. Chem. Chem. Phys.}, 2005, \textbf{7},
  3297--3305\relax
\mciteBstWouldAddEndPuncttrue
\mciteSetBstMidEndSepPunct{\mcitedefaultmidpunct}
{\mcitedefaultendpunct}{\mcitedefaultseppunct}\relax
\EndOfBibitem
\bibitem[Pantazis \emph{et~al.}(2008)Pantazis, Chen, Landis, and
  Neese]{zora_pantazis}
D.~A. Pantazis, X.-Y. Chen, C.~R. Landis and F.~Neese,
  \emph{J.~Chem.~Theory~Comput.}, 2008, \textbf{4}, 908--919\relax
\mciteBstWouldAddEndPuncttrue
\mciteSetBstMidEndSepPunct{\mcitedefaultmidpunct}
{\mcitedefaultendpunct}{\mcitedefaultseppunct}\relax
\EndOfBibitem
\bibitem[Tomasi \emph{et~al.}(2005)Tomasi, Mennucci, and
  Cammi]{pcm-chem-rev-2005}
J.~Tomasi, B.~Mennucci and R.~Cammi, \emph{Chem.~Rev.}, 2005, \textbf{105},
  2999--3094\relax
\mciteBstWouldAddEndPuncttrue
\mciteSetBstMidEndSepPunct{\mcitedefaultmidpunct}
{\mcitedefaultendpunct}{\mcitedefaultseppunct}\relax
\EndOfBibitem
\bibitem[Neese(2022)]{orca-2022}
F.~Neese, \emph{WIREs~Comput.~Mol.~Sci.}, 2022, \textbf{12}, e1606\relax
\mciteBstWouldAddEndPuncttrue
\mciteSetBstMidEndSepPunct{\mcitedefaultmidpunct}
{\mcitedefaultendpunct}{\mcitedefaultseppunct}\relax
\EndOfBibitem
\bibitem[Boguslawski \emph{et~al.}(2021)Boguslawski, Leszczyk, Nowak,
  Brz{\k{e}}k, {\.Z}uchowski, K{\k{e}}dziera, and Tecmer]{pybest-paper}
K.~Boguslawski, A.~Leszczyk, A.~Nowak, F.~Brz{\k{e}}k, P.~S. {\.Z}uchowski,
  D.~K{\k{e}}dziera and P.~Tecmer, \emph{Comput.~Phys.~Commun.}, 2021,
  \textbf{264}, 107933\relax
\mciteBstWouldAddEndPuncttrue
\mciteSetBstMidEndSepPunct{\mcitedefaultmidpunct}
{\mcitedefaultendpunct}{\mcitedefaultseppunct}\relax
\EndOfBibitem
\bibitem[Boguslawski \emph{et~al.}(2024)Boguslawski, Brzęk, Chakraborty,
  Cieślak, Jahani, Leszczyk, Nowak, Sujkowski, Świerczyński, Ahmadkhani,
  Kędziera, Kriebel, Żuchowski, and Tecmer]{pybest-paper-update1-cpc-2024}
K.~Boguslawski, F.~Brzęk, R.~Chakraborty, K.~Cieślak, S.~Jahani, A.~Leszczyk,
  A.~Nowak, E.~Sujkowski, J.~Świerczyński, S.~Ahmadkhani, D.~Kędziera, M.~H.
  Kriebel, P.~S. Żuchowski and P.~Tecmer, \emph{Comput.~Phys.~Commun.}, 2024,
  \textbf{297}, 109049\relax
\mciteBstWouldAddEndPuncttrue
\mciteSetBstMidEndSepPunct{\mcitedefaultmidpunct}
{\mcitedefaultendpunct}{\mcitedefaultseppunct}\relax
\EndOfBibitem
\bibitem[pyb()]{pybest1.3.1-zenodo}
\emph{PyBESTv1.3.1 on zenodo}, \url{https://zenodo.org/records/10069179},
  accessed date: December 12, 2023)\relax
\mciteBstWouldAddEndPuncttrue
\mciteSetBstMidEndSepPunct{\mcitedefaultmidpunct}
{\mcitedefaultendpunct}{\mcitedefaultseppunct}\relax
\EndOfBibitem
\bibitem[pyb()]{pybest-web}
\emph{PyBEST webpage}, \url{http://fizyka.umk.pl/~pybest}, accessed date:
  December 12, 2023)\relax
\mciteBstWouldAddEndPuncttrue
\mciteSetBstMidEndSepPunct{\mcitedefaultmidpunct}
{\mcitedefaultendpunct}{\mcitedefaultseppunct}\relax
\EndOfBibitem
\bibitem[Riplinger and Neese(2013)]{dlpno_neese_1}
C.~Riplinger and F.~Neese, \emph{J.~Chem.~Phys.}, 2013, \textbf{138},
  034106\relax
\mciteBstWouldAddEndPuncttrue
\mciteSetBstMidEndSepPunct{\mcitedefaultmidpunct}
{\mcitedefaultendpunct}{\mcitedefaultseppunct}\relax
\EndOfBibitem
\bibitem[Riplinger \emph{et~al.}(2013)Riplinger, Sandhoefer, Hansen, and
  Neese]{dlpno_neese_2}
C.~Riplinger, B.~Sandhoefer, A.~Hansen and F.~Neese, \emph{J.~Chem.~Phys.},
  2013, \textbf{139}, 134101\relax
\mciteBstWouldAddEndPuncttrue
\mciteSetBstMidEndSepPunct{\mcitedefaultmidpunct}
{\mcitedefaultendpunct}{\mcitedefaultseppunct}\relax
\EndOfBibitem
\bibitem[Cammi(2009)]{cammi2009quantum}
R.~Cammi, \emph{J.~Chem.~Phys.}, 2009, \textbf{131}, 164104\relax
\mciteBstWouldAddEndPuncttrue
\mciteSetBstMidEndSepPunct{\mcitedefaultmidpunct}
{\mcitedefaultendpunct}{\mcitedefaultseppunct}\relax
\EndOfBibitem
\bibitem[Caricato(2011)]{caricato2011ccsd}
M.~Caricato, \emph{J.~Chem.~Phys.}, 2011, \textbf{135}, 074113\relax
\mciteBstWouldAddEndPuncttrue
\mciteSetBstMidEndSepPunct{\mcitedefaultmidpunct}
{\mcitedefaultendpunct}{\mcitedefaultseppunct}\relax
\EndOfBibitem
\bibitem[Barros \emph{et~al.}(2010)Barros, de~Oliveira, Jorge, A., and
  M.]{basis_set_dkh_1}
C.~Barros, P.~de~Oliveira, F.~Jorge, C.~N. A. and C.~M., \emph{Mol.~Phys.},
  2010, \textbf{108}, 1965--1972\relax
\mciteBstWouldAddEndPuncttrue
\mciteSetBstMidEndSepPunct{\mcitedefaultmidpunct}
{\mcitedefaultendpunct}{\mcitedefaultseppunct}\relax
\EndOfBibitem
\bibitem[Jorge \emph{et~al.}(2009)Jorge, Canal~Neto, Camiletti, and
  Machado]{basis_set_dkh_2}
F.~E. Jorge, A.~Canal~Neto, G.~G. Camiletti and S.~F. Machado,
  \emph{J.~Chem.~Phys.}, 2009, \textbf{130}, 064108\relax
\mciteBstWouldAddEndPuncttrue
\mciteSetBstMidEndSepPunct{\mcitedefaultmidpunct}
{\mcitedefaultendpunct}{\mcitedefaultseppunct}\relax
\EndOfBibitem
\bibitem[Douglas and Kroll(1974)]{dkh1}
N.~Douglas and N.~M. Kroll, \emph{Ann.~Phys.}, 1974, \textbf{82}, 89--155\relax
\mciteBstWouldAddEndPuncttrue
\mciteSetBstMidEndSepPunct{\mcitedefaultmidpunct}
{\mcitedefaultendpunct}{\mcitedefaultseppunct}\relax
\EndOfBibitem
\bibitem[Hess(1986)]{dkh2}
B.~A. Hess, \emph{Phys.~Rev.~A}, 1986, \textbf{33}, 3742--3748\relax
\mciteBstWouldAddEndPuncttrue
\mciteSetBstMidEndSepPunct{\mcitedefaultmidpunct}
{\mcitedefaultendpunct}{\mcitedefaultseppunct}\relax
\EndOfBibitem
\bibitem[Reiher and Wolf(2009)]{reiher_book}
M.~Reiher and A.~Wolf, \emph{Relativistic Quantum Chemistry. {T}he Fundamental
  Theory of Molecular Science}, Wiley, 2009\relax
\mciteBstWouldAddEndPuncttrue
\mciteSetBstMidEndSepPunct{\mcitedefaultmidpunct}
{\mcitedefaultendpunct}{\mcitedefaultseppunct}\relax
\EndOfBibitem
\bibitem[{P. Tecmer, K. Boguslawski, D. K{\c{e}}dziera}(2017)]{tecmer2016}
{P. Tecmer, K. Boguslawski, D. K{\c{e}}dziera}, in \emph{Handbook of
  Computational Chemistry}, ed. J.~Leszczy{\'{n}}ski, Springer Netherlands,
  Dordrecht, 2017, vol.~2, pp. 885--926\relax
\mciteBstWouldAddEndPuncttrue
\mciteSetBstMidEndSepPunct{\mcitedefaultmidpunct}
{\mcitedefaultendpunct}{\mcitedefaultseppunct}\relax
\EndOfBibitem
\bibitem[Tecmer \emph{et~al.}(2011)Tecmer, Gomes, Ekstr\"om, and
  Visscher]{pawel1}
P.~Tecmer, A.~S.~P. Gomes, U.~Ekstr\"om and L.~Visscher,
  \emph{Phys.~Chem.~Chem.~Phys.}, 2011, \textbf{13}, 6249--6259\relax
\mciteBstWouldAddEndPuncttrue
\mciteSetBstMidEndSepPunct{\mcitedefaultmidpunct}
{\mcitedefaultendpunct}{\mcitedefaultseppunct}\relax
\EndOfBibitem
\bibitem[Berman \emph{et~al.}(2000)Berman, Westbrook, Feng, Gilliland, Bhat,
  Weissig, Shindyalov, and Bourne]{pdb}
H.~M. Berman, J.~Westbrook, Z.~Feng, G.~Gilliland, T.~N. Bhat, H.~Weissig,
  I.~N. Shindyalov and P.~E. Bourne, \emph{Nucleic Acids Res.}, 2000,
  \textbf{28}, 235--242\relax
\mciteBstWouldAddEndPuncttrue
\mciteSetBstMidEndSepPunct{\mcitedefaultmidpunct}
{\mcitedefaultendpunct}{\mcitedefaultseppunct}\relax
\EndOfBibitem
\end{mcitethebibliography}
\bibliographystyle{rsc} 

\end{document}